\documentclass[twocolumn,epjc3]{svjour3}  

\RequirePackage{graphicx}
\RequirePackage{mathptmx}      
\RequirePackage{flushend}
\RequirePackage{color}
\usepackage{amsmath}
\usepackage{hyperref}
\usepackage{color}
\usepackage[dvipsnames]{xcolor}
\usepackage[section]{placeins}
\usepackage[normalem]{ulem}
\usepackage{appendix}

\newcommand{\gevcc}{GeV/$c^2$}

\newcommand{\gev}{GeV}
\newcommand{\tev}{TeV}
\newcommand{\agev}{$A$ GeV}

\newcommand{\sqrtsnn}{$\sqrt{s_\textrm{NN}}$}

\newcommand{\mt}{m_{t}}
\newcommand{\mtmt}{m_{t}^{2}}
\newcommand{\TOne}{T_{1}}
\newcommand{\TTwo}{T_{2}}
\newcommand{\Ei}{E_{i}}
\newcommand{\Ef}{E_{f}}
\newcommand{\Epi}{E_{\pi}}

\newcommand{\Tp}{T_{p}}
\newcommand{\pini}{p_{i}}
\newcommand{\pf}{p_{f}}
\newcommand{\Vc}{V_{C}}
\newcommand{\Mp}{m_{p}}
\newcommand{\Mpi}{m_{\pi}}
\newcommand{\Veff}{V_{\textrm{eff}}}
\newcommand{\ycm}{y_{\textrm{c.m.}}}

\newcommand{\pim}{$\pi^{-}$}
\newcommand{\pip}{$\pi^{+}$}
\newcommand{\Jeff}{J_{\textrm{eff}}}

\DeclareMathOperator\erf{erf}

\journalname{Eur. Phys. J. A}

\begin{document}

\title{Impact of the Coulomb field on charged-pion spectra in few-GeV heavy-ion collisions$^{*}$}

\author{HADES collaboration\\[10bp]
J.~Adamczewski-Musch$^{5}$, O.~Arnold$^{11,10}$, C.~Behnke$^{9}$, A.~Belounnas$^{17}$,
A.~Belyaev$^{8}$, J.C.~Berger-Chen$^{11,10}$, A.~Blanco$^{2}$, C.~Blume$^{9}$, M.~B\"{o}hmer$^{11}$,
P.~Bordalo$^{2}$
% ...
}

\institute{
\mbox{} \\[-8bp]
\mbox{$^{1}$Institute of Physics, Slovak Academy of Sciences, 84228~Bratislava, Slovakia}\\
\mbox{$^{2}$LIP-Laborat\'{o}rio de Instrumenta\c{c}\~{a}o e F\'{\i}sica Experimental de Part\'{\i}culas , 3004-516~Coimbra, Portugal}\\
\mbox{$^{3}$Institute of Nuclear Physics, Polish Academy of Sciences, 31342~Krak\'{o}w, Poland}\\
\mbox{$^{4}$Smoluchowski Institute of Physics, Jagiellonian University of Cracow, 30-059~Krak\'{o}w, Poland}\\
\mbox{$^{5}$GSI Helmholtzzentrum f\"{u}r Schwerionenforschung GmbH, 64291~Darmstadt, Germany}\\
\mbox{$^{6}$Technische Universit\"{a}t Darmstadt, 64289~Darmstadt, Germany}\\
\mbox{$^{7}$Institut f\"{u}r Strahlenphysik, Helmholtz-Zentrum Dresden-Rossendorf, 01314~Dresden, Germany}\\
\mbox{$^{8}$Joint Institute of Nuclear Research, 141980~Dubna, Russia}\\
\mbox{$^{9}$Institut f\"{u}r Kernphysik, Goethe-Universit\"{a}t, 60438 ~Frankfurt, Germany}\\
\mbox{$^{10}$Excellence Cluster 'Origin and Structure of the Universe' , 85748~Garching, Germany}\\
\mbox{$^{11}$Physik Department E62, Technische Universit\"{a}t M\"{u}nchen, 85748~Garching, Germany}\\
\mbox{$^{12}$II.Physikalisches Institut, Justus Liebig Universit\"{a}t Giessen, 35392~Giessen, Germany}\\
\mbox{$^{13}$Institute for Nuclear Research, Russian Academy of Science, 117312~Moscow, Russia}\\
\mbox{$^{14}$Institute of Theoretical and Experimental Physics, 117218~Moscow, Russia}\\
\mbox{$^{15}$National Research Nuclear University MEPhI (Moscow Engineering Physics Institute), 115409~Moscow, Russia}\\
\mbox{$^{16}$Department of Physics, University of Cyprus, 1678~Nicosia, Cyprus}\\
\mbox{$^{17}$Laboratoire de Physique des 2 infinis Ir\`{e}ne Joliot-Curie, Universit\'{e} Paris-Saclay, CNRS-IN2P3, F-91405~Orsay, France}\\
\mbox{$^{18}$Nuclear Physics Institute, The Czech Academy of Sciences, 25068~Rez, Czech Republic}\\
\mbox{$^{19}$LabCAF. F. F\'{\i}sica, Universidad de Santiago de Compostela, 15706~Santiago de Compostela, Spain}\\
\mbox{$^{20}$Uniwersytet Warszawski, Wydzia{\l} Fizyki, Instytut Fizyki Do\'{s}wiadczalnej, 02-093~Warszawa, Poland}\\
\mbox{$^{21}$Dipartimento di Fisica and INFN, Universit\`{a} di Torino, 10125~Torino, Italy}\\
\\
\mbox{$^{*}$ The preparation of this manuscript was completed before February 24, 2022}\\
\mbox{$^{a}$ also at Coimbra Polytechnic - ISEC, ~Coimbra, Portugal}\\
\mbox{$^{b}$ also at Technische Universit\"{a}t Dresden, 01062~Dresden, Germany}\\
\mbox{$^{c}$ also at Frederick University, 1036~Nicosia, Cyprus}\\
\mbox{$^{\dag}$ deceased} \\
\mbox{e-mail: hades-info@gsi.de}\\
}

\authorrunning{J.~Adamczewski-Musch {\em et al.}}
\titlerunning{Impact of the Coulomb field on charged-pion spectra in few-GeV heavy-ion collisions}

\date{\today}


\maketitle

\begin{abstract}
In nuclear collisions the incident protons generate a Coulomb field which acts on produced charged particles. The impact of these interactions on charged-pion transverse-mass and rapidity
spectra, as well as on pion-pion momentum correlations is investigated in Au+Au collisions at \sqrtsnn\ = 2.4~GeV.  We show that the low-$m_t$ region ($m_t < 0.2$~ \gevcc)
can be well described with a Coulomb-modified Boltzmann distribution that also takes changes of the Coulomb field during the expansion of the fireball into account. The observed centrality
dependence of the fitted mean Coulomb potential energy deviates strongly from a $A_{\textrm{part}}^{2/3}$ scaling, indicating that, next to the fireball, the non-interacting charged spectators have
to be taken into account. For the most central collisions, the Coulomb modifications of the HBT source radii are found to be consistent with the potential extracted from the single-pion
transverse-mass distributions.  This finding suggests that the region of homogeneity obtained from two-pion correlations coincides with the region in which the pions freeze-out.  Using the
inferred mean-square radius of the charge distribution at freeze-out, we have deduced a baryon density, in fair agreement with values obtained from statistical hadronization model fits to the particle yields.

\end{abstract}

\section{Introduction}

The fireball produced in a relativistic heavy-ion collision has a net positive electric charge due to the protons from both target and projectile nuclei.  Charged particles emitted from the expanding fireball are subjected
to the long-range Coulomb interaction caused by this electric charge resulting in distortions of their emission spectra.  Such modifications can be observed in the spectra of charged hadrons, ideally by comparing
the effects on oppositely charged states.  Positively and negatively charged pions are consequently ideal probes of the Coulomb effects: they have the same mass but opposite charge, and are produced copiously
with comparable yields already in few-GeV collisions.  The force exerted by the electric field accelerates the $\pi^+$ and decelerates the $\pi^-$ resulting in a change of their kinetic energies.
The corresponding average energy kick is determined by the charge distribution of the incident protons, which create the Coulomb field, as well as by the distribution of pion emission points, i.e.\ the pion source.
Investigating the Coulomb effects has the potential to reveal information on the characteristics of the pion source at freeze-out.
An increase of the $\pi^-/\pi^+$ yield ratio at low momenta was first observed in early Bevalac experiments \cite{Benenson:1979,Nagamiya:1981,Sullivan:1982}
and was quickly interpreted as being related to the Coulomb field \cite{Libbrecht:1979tu,Gyulassy:1980xb}.  Indeed, these effects turned out to be ubiquitous in heavy-ion collisions, ranging from SIS
beam energies \cite{Muentz:1998odc,Wagner:1997sa,Hong:2005at}, to AGS \cite{Barrette:1999ry,Cebra:2014sx}, SPS \cite{Boggild:1996sy}, and even up to RHIC energies, as discussed in an
extensive report \cite{Cebra:2014sx}.  The early theoretical work done to interpret the Bevalac data \cite{Libbrecht:1979tu,Gyulassy:1980xb} was taken up again as the SIS, AGS, and SPS results became available \cite{BaoanLi:1995bb,Baym:1996wk,Barz:1997su,Ayala:1997aa,Ayala:1998gu,Rybicki:2006qm,Ozvenchuk:2019cve}.  In particular at low bombarding energies, the $\pi^{-}/\pi^{+}$ ratio is expected to provide information on the
symmetry term of the nuclear equation of state \cite{Estee2021}, thus requiring the Coulomb effect to be under full control \cite{Stone:2022mho}.  Finally, the Coulomb field also influences the distributions of
relative momenta used in two-pion interferometry \cite{Zajc:1984,Chacon:1988,Chacon:1991,Christie:1992} and this must be taken into account to properly interpret the measured HBT radii of charged-particle
sources \cite{Baym:1996wk,Pratt:1986ev,Barz:1996gr,Barz:1998ce,Maj:2009ue,Adamczewski-Musch:2018owj,Adamczewski-Musch:2019dlf}.

The High Acceptance Di-electron Spectrometer (HADES) experiment \cite{Agakichiev:2009} at the SIS18 accelerator has extensively studied particle production in 1.23 \agev\ Au+Au collisions, equivalent to a
center-of-mass energy of \sqrtsnn\ = 2.4~GeV.  In particular, charged-pion spectra were measured with very high statistics and the pion phase-space distribution could be reconstructed with good accuracy \cite{Adamczewski-Musch:2020vrg},
limited mostly by the systematic uncertainties of the detector efficiency and acceptance of order 5--10\%.  Especially, the extrapolation of the measured differential particle yields into the low-$m_t$ region, not fully covered by HADES,
is a potential source of systematic uncertainties. 
In the past, Boltzmann distributions were often used for this extrapolation, neglecting the spectral distortions caused by the Coulomb interaction.  One motivation for the analysis presented here was to provide
a more accurate parameterization of pion distributions at low $m_t$ (or low $p_t$), allowing to reduce systematic uncertainties due to the extrapolation of measured yields into this region.   This is particularly important
owing to the fact that the geometry of the HADES detector and its toroidal magnetic field lead to a somewhat different low-momentum cut-off of the $\pi^-$, bent to large polar angles, and the $\pi^+$, bent towards
the beam axis \cite{Agakichiev:2009,Adamczewski-Musch:2020vrg}.

This paper is organized as follows.  In section~2 we introduce the formalism used to describe the Coulomb effect on charged-particle spectra and we validate our fit procedure by simulations.  In section~3, we present
an improved analysis of the charged-pion spectra measured in central and semi-peripheral Au+Au collisions at \sqrtsnn\ = 2.4~GeV.   The extracted values of the average Coulomb potential energy $V_C$ as well as
the resulting source radius are discussed as a function of the collision centrality.   We then compare these results with estimates of the Coulomb potential energy derived from the HBT radii of the measured like-sign
pion-pion momentum correlations.  From the pion source volume, we extract a baryon density at freeze-out which is compared with the Statistical Hadronization Model (SHM) fits to the particle yields in the same event sample.
Using published data from other experiments, we then present the excitation function of $V_C$ from 1 to 10 \agev\ kinetic beam energy.  In section~4, finally, we summarize our findings and finish with a brief outlook.  

\section{Methodology}
\subsection{Coulomb field acting on charged pions}

In a central collision at \sqrtsnn\ = 2.4~GeV, the colliding Au nuclei are stopped in a fireball which afterwards expands radially with a velocity of about $\langle \beta \rangle \approx$ 0.3 -- 0.4 \cite{Reisdorf2010,Schuldes2016},
with the produced particles moving away in the Coulomb field of the positive net charge of the fireball.  The influence of the Coulomb field is most noticeable in the velocity or momentum distributions of low mass charged
particles, in particular pions:  $\pi^{+}$ are sped up and $\pi^{-}$ are slowed down, causing substantial modification of their differential yields.  Due to the resulting reshuffling, at low center-of-mass momenta,
the $\pi^{+}$ yield is reduced whereas the $\pi^{-}$ yield is increased relative to the uncharged pions.  In collisions in which the nuclei are not fully stopped their longitudinal motion as well as possible charged spectators
need to be taken into account.

The nucleus-nucleus center-of-mass energy in the final state $\Ef$ of pions can be expressed in terms of their initial energy $\Ei$, i.e.~the energy corrected for the presence of the electrostatic field, and the Coulomb
potential energy $\Vc$ as
\begin{equation}
E^{\pm}_f(\pf) = \Ei(\pini) \pm \Vc \;,
\label{eq-coul}
\end{equation}

\noindent
where $p_f$ and $p_i$ are the corresponding final and initial pion center-of-mass momenta, and the $\pm$ sign corresponds to the different pion charges.  As the total charge of the fireball is due to the incoming
proton charges, $V_C$ is a positive quantity.  In this simple static picture, the Coulomb potential energy leads to a shift of the pion total energy by an amount $+V_C (-V_C)$ for positively (negatively) charged pions.
However, as discussed in \cite{Barz:1997su}, the expansion of the charged fireball causes an attenuation of the Coulomb effect. Indeed, the central electric field felt by a given pion is only produced by
those charges that are slower than this pion, i.e.~the charged matter shell that has overtaken the pion in the outward expansion does not contribute.  For a complete description of the expansion, a fully dynamic calculation
would have to be used, e.g.~within a hadronic transport model allowing for different freeze-out times and loci depending on the charge and momentum of the pions.  Here, we follow instead a more data-driven ansatz similar
to the one applied in \cite{Cebra:2014sx,Barz:1997su}, where all pions are assumed to freeze-out at the same time, and parameterize the attenuation of the electrostatic force by replacing $\Vc$ with an effective
potential $\Veff$ expressed as a function of the pion kinetic energy $\Epi - m_{\pi}$.\footnote{In a classical approximation, the average kinetic energy of thermal protons is $\langle E^p_{\textrm{kin}} \rangle = 3/2 \, k \, T_p$,
e.g. for an effective proton temperature $T_p = 130$~MeV, $\langle E^p_{kin} \rangle = 195$~MeV.    The pions of the same velocity, i.e.~with a kinetic energy $E^{\pi}_{kin} = m_{\pi}/m_p \times 195 = 29$~MeV,
experience a Coulomb field attenuated by a factor $1 - e^{-1.5} \simeq 3/4 (3/5)$ in the 2D (3D) case.}  Defining $x = \sqrt{(E_{\pi}/m_{\pi} - 1) \, m_p/T_p}$, which is a measure for the relative velocities of pions and protons,
and integrating over the proton velocity distribution, the authors of \cite{Barz:1997su} found

\begin{equation}
\Veff \;\; =  \;\;
\begin{cases}
 \; \Vc \, \left(1-e^{-x^2} \right) & \textrm{for 2D expansion} ,\\
 \; \Vc \, \left( \erf(x) - (2/\sqrt{\pi}) \, x \, e^{-x^2} \right) & \textrm{for 3D expansion} , 
\end{cases}
\label{eq-Veff-cases}
\end{equation}
~\\
\noindent
where $m_p$ is the proton mass, $\Tp$ is the inverse slope parameter from a fit to the proton $\mt$ distribution, and \textrm{$\erf(x)$} is the error function.
The 2D case corresponds to a cylindrical geometry with a transverse expansion (i.e.~for boost-invariant systems, relevant at high collision energies) and the 3D case
stands for a spherical expansion.  The latter one is more appropriate for the low beam energies at which HADES operates.   The rationale behind Eq.~\eqref{eq-Veff-cases}
is the following:  the protons provide the dominant part of the fireball net charge and the attenuation term is given by the fraction of protons that are slower than
a given outgoing pion.  Note that the proton energy distribution can be described alternatively by a blast-wave fit, using e.g.~Siemens-Rasmussen \cite{SiemRas:1979aa},
although in that case no closed formula is available for $V_{\textrm{eff}}$ in Eq.~\eqref{eq-Veff-cases}.   The integration over the proton velocity distribution would have
to be carried out numerically, increasing massively the cost for repeatedly computing $V_{\textrm{eff}}$ in the iterative fitting procedure.  Ultimately, the attenuation
of the Coulomb field as a function of pion energy is realized in our fit function by replacing the parameter $\Vc$ in Eq.~\eqref{eq-coul} with $\Veff$ of
Eq.~\eqref{eq-Veff-cases}, and $V_{\textrm{eff}}$ taking the sign of $V_C$.

To extrapolate pion spectra into phase-space regions not covered by experiment, it is customary to use a relativistic Boltzmann distribution adjusted to the data.
In Ref.~\cite{Adamczewski-Musch:2020vrg}, we had shown that a proper description of the transverse-mass spectrum requires in fact the sum of two Boltzmann distributions,
expressed as

\begin{equation}
 \frac{d^{2}N^{\pm}}{d\mt dy} =  A \, \mtmt \left( f e^{-E/ \TOne} + (1-f)e^{-E/ \TTwo} \right) \,,
\label{eq-fit-mt}
\end{equation}
~\\
\noindent
where $E$ is the total energy of the pion in the center of mass,\footnote{Ignoring the Coulomb force, $E = E_f = E_i$.} $A$ is the normalization, $\TOne$ and $\TTwo$ are
the slopes of the two spectral components, and $f$ and $1-f$ are their corresponding fractional amplitudes ($0 < f \leq 1$).   Following Ref.~\cite{Barz:1997su},
we insert the initial total pion energy $\Ei$ into Eq.~\eqref{eq-fit-mt} and, with the help of Eq.~\eqref{eq-coul}, express it as a function of the observed
final energy $\Ef$ to obtain

\begin{multline}
 \frac{d^{2}N^{\pm}}{d\mt dy} = A \, \mtmt \left( f \,e^{-(\Ef \mp \Veff)/ \TOne} + (1-f) \,e^{-(\Ef \mp \Veff)/ \TTwo} \right) \\
 \times J \times \Jeff \,.
\label{eq-fit-mt-Veff}
\end{multline}
~\\
\noindent
This distribution contains the full Jacobian $J \times J_{\mathrm{eff}}$ of the transformation from "initial" to "final" kinematic variables. 
It consists of two parts: the first factor $J$ corresponds to the Jacobian proposed in Ref.~\cite{Baym:1996wk} for a constant Coulomb potential energy, namely

\begin{equation}
 J = \frac{\Ei \, \pini}{\Ef \, \pf} = \frac{(\Ef \mp \Veff) \, \sqrt{(\Ef \mp \Veff)^2 - \Mpi^2}}{\Ef \, \sqrt{\Ef^2 - \Mpi^2}} \,. 
 \label{eq-fit-jac}
 \end{equation}
~\\
\noindent
The second factor $J_{\mathrm{eff}}$ results from the explicit dependence of $\Veff$ on the pion kinetic energy introduced via Eq.~\eqref{eq-Veff-cases}. As shown in Appendix A, it is expressed as
 
 \begin{equation} 
 \Jeff \;\;\; =  \;\;\;
\begin{cases}
 \;\;\; 1 \mp \frac{\Vc\;\Mp}{\Mpi\;\Tp} \; e^{-x^2} & \textrm{for 2D expansion} ,\\
 \;\;\; 1 \mp \frac{2}{\sqrt{\pi}} \;\; \frac{\Vc \;\Mp}{\Mpi\;\Tp} \; x \; e^{-x^2} & \textrm{for 3D expansion} , 
\end{cases}
\label{eq-fit-jaceff}
\end{equation}
~\\
\noindent
with $x$ as defined above for Eq.~\eqref{eq-Veff-cases}.

Note that this second factor has been omitted in previous work, e.g.~in \cite{Cebra:2014sx,Barz:1997su} and is here applied for the first time in the analysis.  As demonstrated on simulated pion spectra in the next subsection,
we find that the extracted values for the Coulomb potential energy would come out almost twice larger if this factor was not included. 

\subsection{Effect on the kinematic distribution of charged pions}

\begin{figure}[ht]
    \centering
	\includegraphics[width=0.49\linewidth]{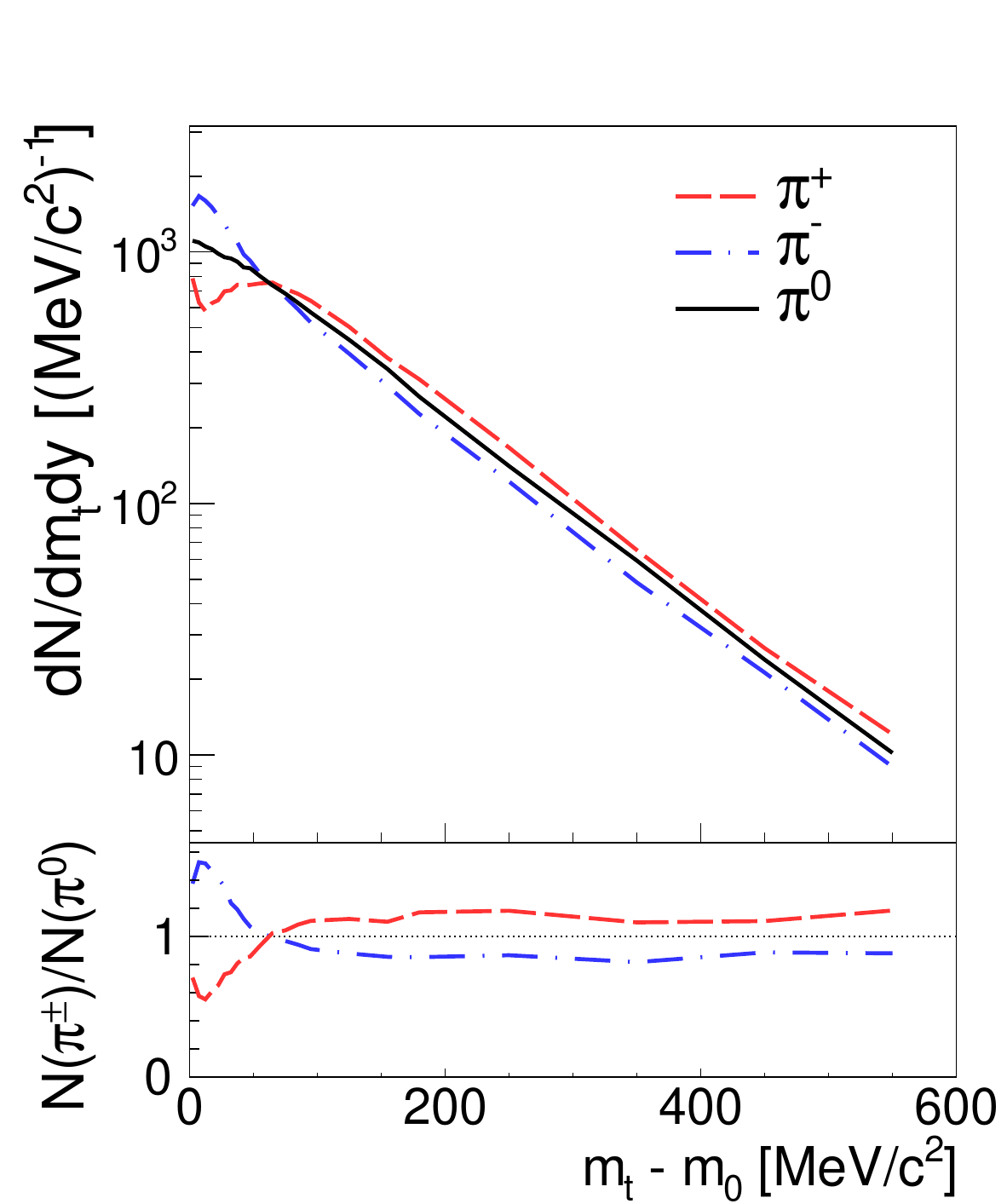}
	\includegraphics[width=0.49\linewidth]{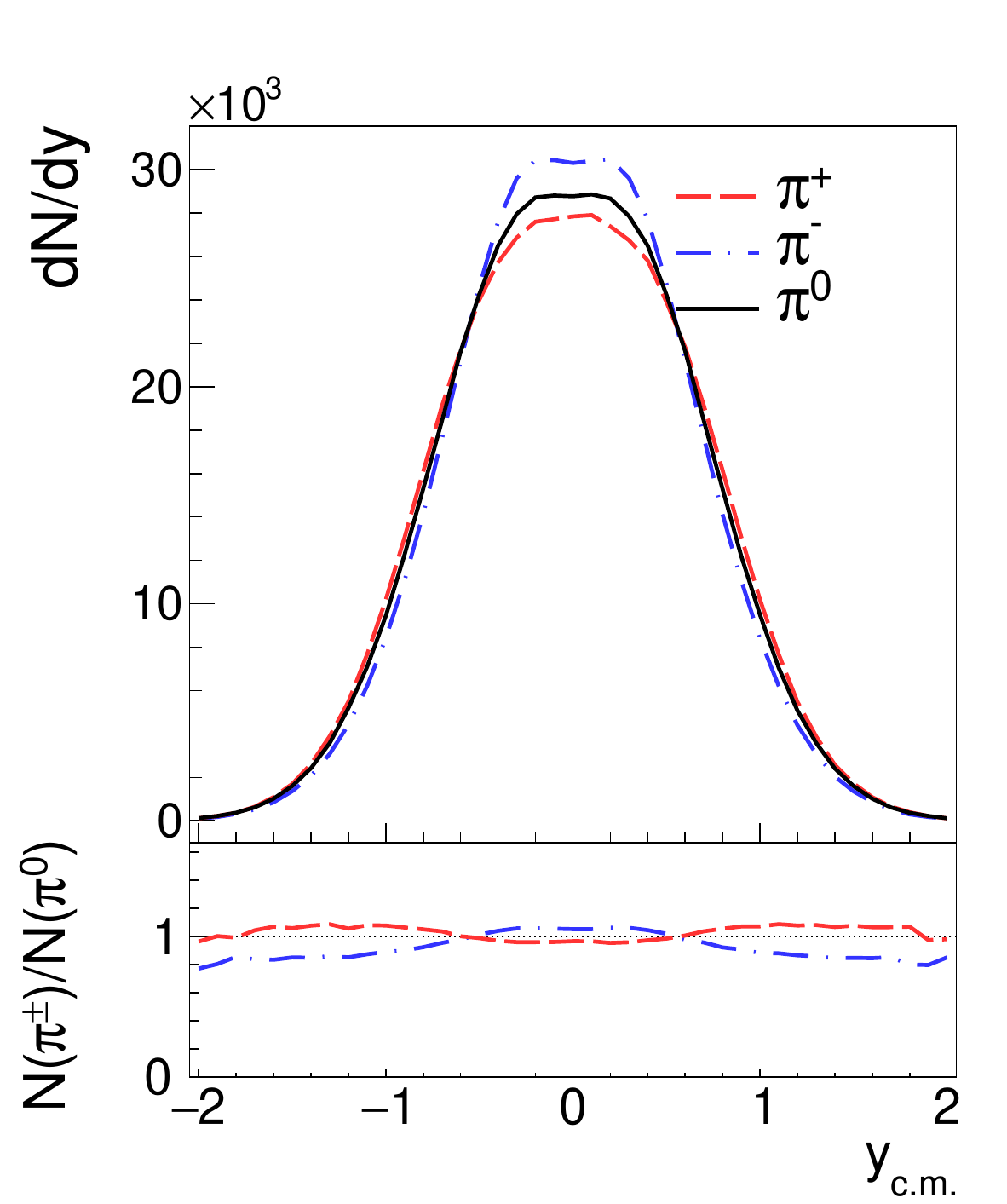}
	\caption{Left: Simulated spectral distributions at mid-rapidity of $\pi^-$ (blue) and $\pi^+$ (red) as a function of $m_t-m_0$, compared to the pure Boltzmann case ($\pi^0$, black); the lower panel shows the ratios
   of charged pions with respect to $\pi^{0}$.  
   Right:  Simulated center-of-mass rapidity distributions of $\pi^-$ (blue), $\pi^+$ (red) and $\pi^0$ (black);
   the lower panel shows the ratios of charged pions with respect to $\pi^{0}$. The simulation was done with the Pluto event generator using a common set of parameters for all pion charges; see text for details.}
   \label{fig-effect}  
\end{figure}

In order to illustrate the effect of a Coulomb field on the kinematic distributions of charged pions, we have used the event generator Pluto \cite{Pluto:2009aa} to simulate thermal spectra of all three pion species, $\pi^+$ $\pi^0$,
and $\pi^-$. To do that, the Pluto code was modified to include the Coulomb potential energy, basically by implementing Eqs.~\eqref{eq-fit-mt-Veff} -- \eqref{eq-fit-jaceff} in the energy sampling routine.  In the simulation,
we have set the parameters to $T_1 = 50$~MeV, $T_2 = 90$~MeV, $f = 0.95$, $V_c = 15$~MeV, $T_p = 130$~MeV, and used as well angular distribution coefficients set to $A_2$ = 0.54/0.60/0.66 for $\pi^{+}/\pi^0/\pi^{-}$,
which are typical values observed for the energy regime where HADES operates \cite{Adamczewski-Musch:2020vrg,Schuldes2016}. 
The left panel of Fig.~\ref{fig-effect} depicts the simulated distributions around mid-rapidity ($0.74 \pm 0.05$) for $\pi^-$ and $\pi^+$ as a function of $m_t-m_0$, compared to the pure $\pi^{0}$ case, i.e.~without the
effect of the Coulomb field, naturally realized by neutral pions.  In order to focus on the Coulomb effect, isospin related differences were ignored, i.e.\ the three pion flavors were simulated with equal multiplicities.
Fluctuations at the higher $m_t$ values are caused by the limited event statistics.  The lower panel depicts the ratios of charged pions with respect to the $\pi^{0}$ distribution.  The effect of the Coulomb potential is most
prominent at very low $m_t-m_0$ leading to an enhancement of the $\pi^-$ and a depletion of the $\pi^+$ by almost 50\%.  At high $m_t-m_0$, the effect is opposite for both $\pi^-$ and $\pi^+$, amounting to about 15\%.
Note that, for the determination of the total pion yield from measured data, the low part of the transverse-mass spectrum is of particular interest: it holds a substantial fraction of the yield, while often requiring extrapolation
to correct for incomplete detector acceptance.
 
The right panel of Fig.~\ref{fig-effect} displays the simulated $\pi^-$ (blue) and $\pi^+$ (red) center-of-mass rapidity distributions $dN/dy$, again compared to the $\pi^0$ case (black);  the lower panel depicts the ratios
relative to the $\pi^{0}$ distribution.  At mid-rapidity, the effect of the Coulomb field on the total pion yields amounts to about 5\%.  In the tails, however, the relative differences are larger, with the Coulomb force leading
to a narrowing of the $\pi^-$ distribution and a broadening of the $\pi^+$ one relative to $\pi^{0}$.

\subsection{Test of the fit procedure}

From data, the parameter $\Vc$ can be obtained most robustly by adjusting Eq.~\eqref{eq-fit-mt-Veff} simultaneously to both the measured \pip\ and \pim~ transverse-mass
(transverse-momentum) spectra.  Before turning to measured data, however, we have cross checked the performance of our fitting procedure using a Monte Carlo simulation.
Employing the event generator Pluto \cite{Pluto:2009aa}, we have set up a pion source according to Eq.~\eqref{eq-fit-mt-Veff} with the realistic values of the
parameters $\TOne$, $\TTwo$, $f$, and $\Vc$, as given before.  In this simulation we have assumed a spherical expansion of the fireball, as described
by Eq.~\eqref{eq-Veff-cases} for the 3D case.

\begin{figure}[ht]\	\centering
	\includegraphics[width=0.8\linewidth]{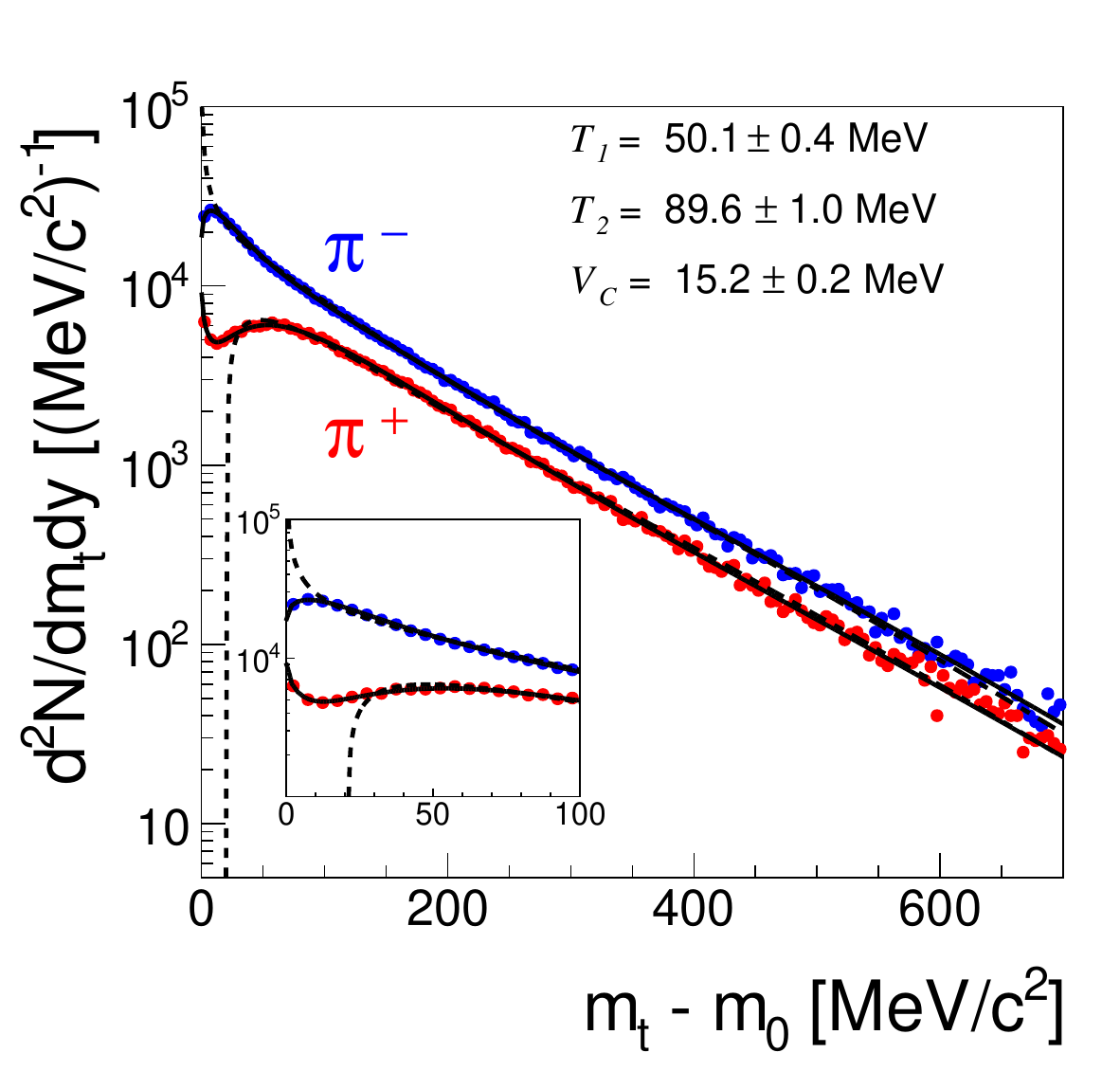}
	\caption{Pion spectra at mid-rapidity simulated with the event generator Pluto (using $A_{-}/A_{+} = 2$, $T_1 = 50$~MeV, $T_2 = 90$~MeV, $f_{\pm} = 0.95$,
          $V_C = 15$~MeV).  Solid curves represent a simultaneous fit to the \pip and \pim spectra including a $V_{\mathrm{eff}}$ attenuated by the 3D fireball expansion;
          dashed curves correspond to a fit done with the non-attenuated, i.e. full potential $V_C$ (see text).  The parameters obtained from the fit with
          $V_{\mathrm{eff}}$ and their statistical error bars are listed in the upper box; the insert shows a close-up of the low-$\mt$ region.}
	\label{fig-pluto-fit}  
\end{figure}	

Figure~\ref{fig-pluto-fit} shows the simulated charged-pion $dN/dm_t$ spectra at mid-rapidity, together with fits using the Coulomb-modified two-component Boltzmann
distribution of Eq.~\eqref{eq-fit-mt-Veff}.  As already pointed out, the influence of the Coulomb field on the charged pions manifests itself mostly at low $m_t - m_0$.
From the figure one can see that the values of the fitted parameters agree within statistical errors with the input, validating our procedure.  In order to illustrate the
impact of using an attenuated potential, we present two cases: first, using a $\Veff$ to account for 3D expansion of the fireball (Eq.~\eqref{eq-fit-mt-Veff}) and, second,
ignoring the expansion, i.e.\ setting $\Veff = \Vc$ and $\Jeff = 1$.  The two fits are displayed in Fig.~\ref{fig-pluto-fit} as solid and dashed colored curves, respectively.
As the inserted close-up shows, the two fit curves differ substantially at very low $m_t$, giving also largely different values of the extracted Coulomb potential energy
$V_C = 15.2$~MeV ($\chi^2/\mathrm{ndf} = 1.07$) compared to $V_C = 20.9$~MeV ($\chi^2/\mathrm{ndf} = 3.08$).  We investigated as well the fitting scheme proposed
in Ref~\cite{Cebra:2014sx}, realized by Eq.~\eqref{eq-fit-mt-Veff} with $J_\mathrm{eff} = 1$, resulting in $V_C = 32.8$~MeV ($\chi^2/\mathrm{ndf} = 1.38$),
i.e.\ more than double the value used to simulate the events.

\subsection{Role of the spectator protons}

\begin{figure}[ht]
	\centering
	\includegraphics[width=0.7\linewidth]{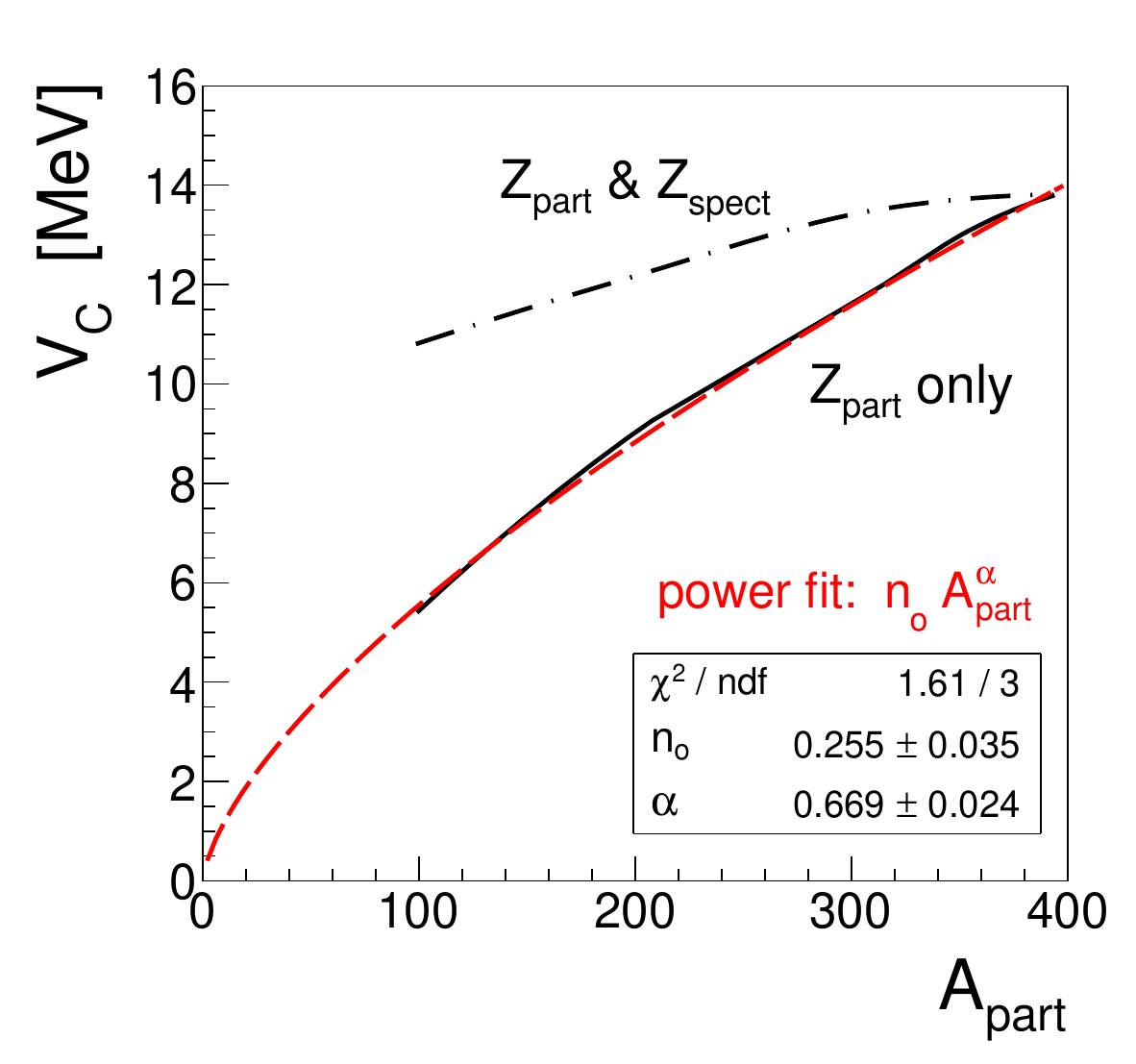}
	\caption{Effect of the contributing spectator charge $Z_{\mathrm{spect}}$ on the adjusted parameter $V_C$.  Shown is $V_C$ as a function of $A_{\mathrm{part}}$
          obtained from fits of Eq.~\eqref{eq-fit-mt-Veff} to mid-rapidity pion spectra simulated with and without spectator charge included (see text for details).
          Ignoring the spectator charge, a power fit to $V_C(A_{\mathrm{part}})$ yields the expected scaling $V_C \propto A_{\mathrm{part}}^{ 2/3}$ (red dashed line).}
    \label{fig-Simul-Vc-Apart}  
\end{figure}

In the picture of an expanding charged fireball, only the total charge of the participant nucleons contributes to the Coulomb effect.  The implicit assumption is that the
Coulomb field of the spectator protons is small,  and that the latter move away fast enough to be neglected. As a consequence, $V_C$ would scale with $A_{\mathrm{part}}$
like the ratio of volume over radius of the charge distribution, i.e.\ $V_C \propto Z_{\mathrm{part}}/R \propto A_{\mathrm{part}}^{2/3}$.  However, in order to estimate the
influence of spectator charges, we have implemented in our simulation the three-source ansatz of Gyulassy and Kauffmann \cite{Gyulassy:1980xb} by defining in Pluto
a charged fireball at rest in the center of mass and two receding charged spectators.  In this scheme, the pions are emitted from the participant zone only, but are
affected by the combined Coulomb field of all three charges. By fitting Eq.~\eqref{eq-fit-mt-Veff} to the generated pion spectra at mid-rapidity, the evolution with
$A_{\mathrm{part}}$ of the resulting effective Coulomb potential energy can be determined.  The fitted parameter $V_C$ is shown in Fig.~\ref{fig-Simul-Vc-Apart} for
calculations performed with and without the spectators included.  While still not being dynamic, this simulation illustrates the trend of $V_C$ with $A_{\mathrm{part}}$,
suggesting that, while spectator contributions are indeed small in central collisions, already in semi-peripheral and more so in peripheral collisions, the Coulomb
potential energy deviates from the simple, central-source $A_{\mathrm{part}}^{2/3}$ scaling.   Evidently, the interpretation of Coulomb effects in peripheral events
or off mid-rapidity will be more demanding than for central events.  The influence of spectator charges on pion spectra has also been investigated
in Refs.~\cite{Rybicki:2006qm,Ozvenchuk:2019cve}, but for SPS energies only.

\color{Black}
\section{Results and Discussion}
\subsection{Effect of the Coulomb field in Au+Au collisions}

\begin{figure}[htb]
	\centering
	\includegraphics[width=0.49\linewidth]{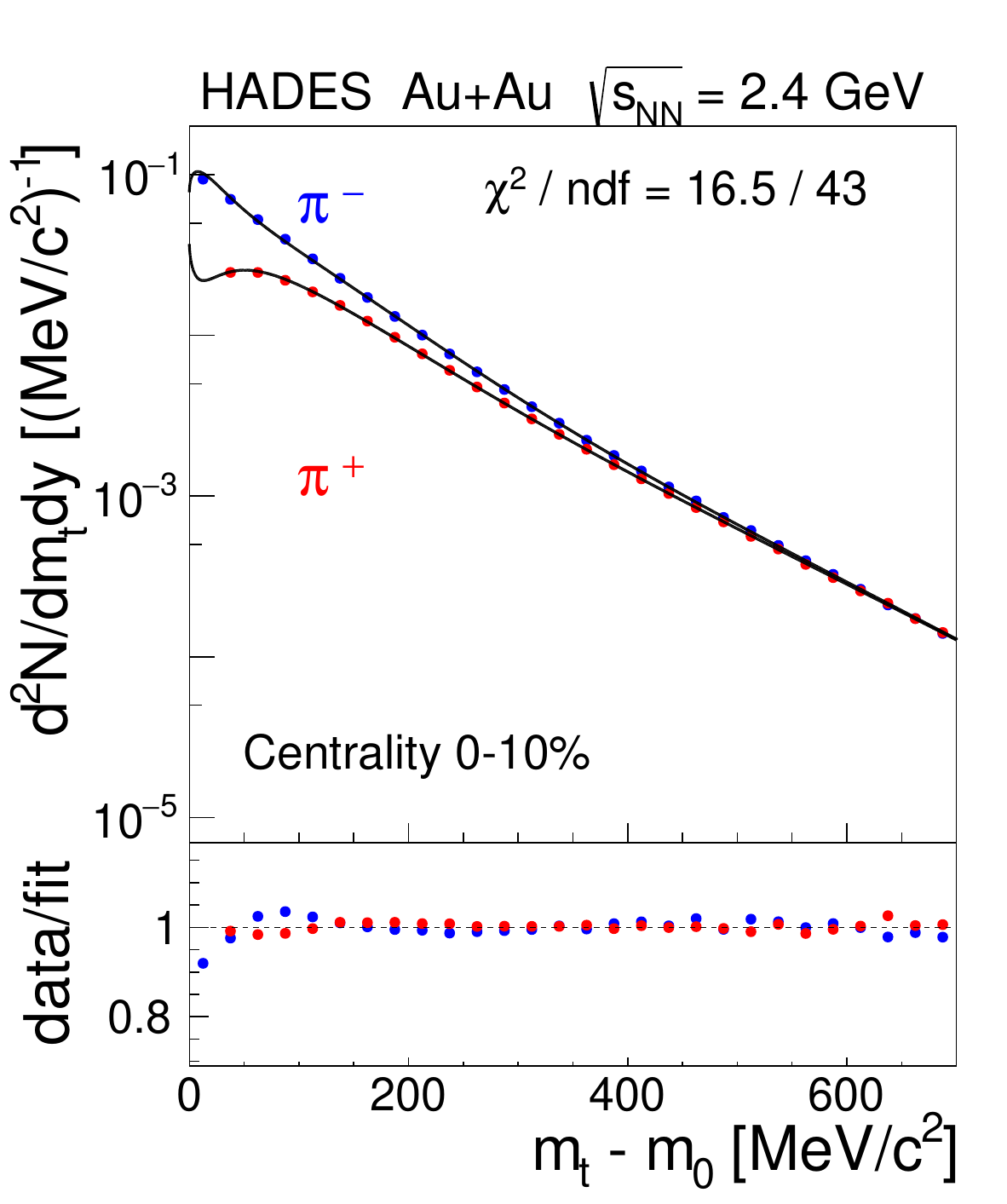}
	\includegraphics[width=0.49\linewidth]{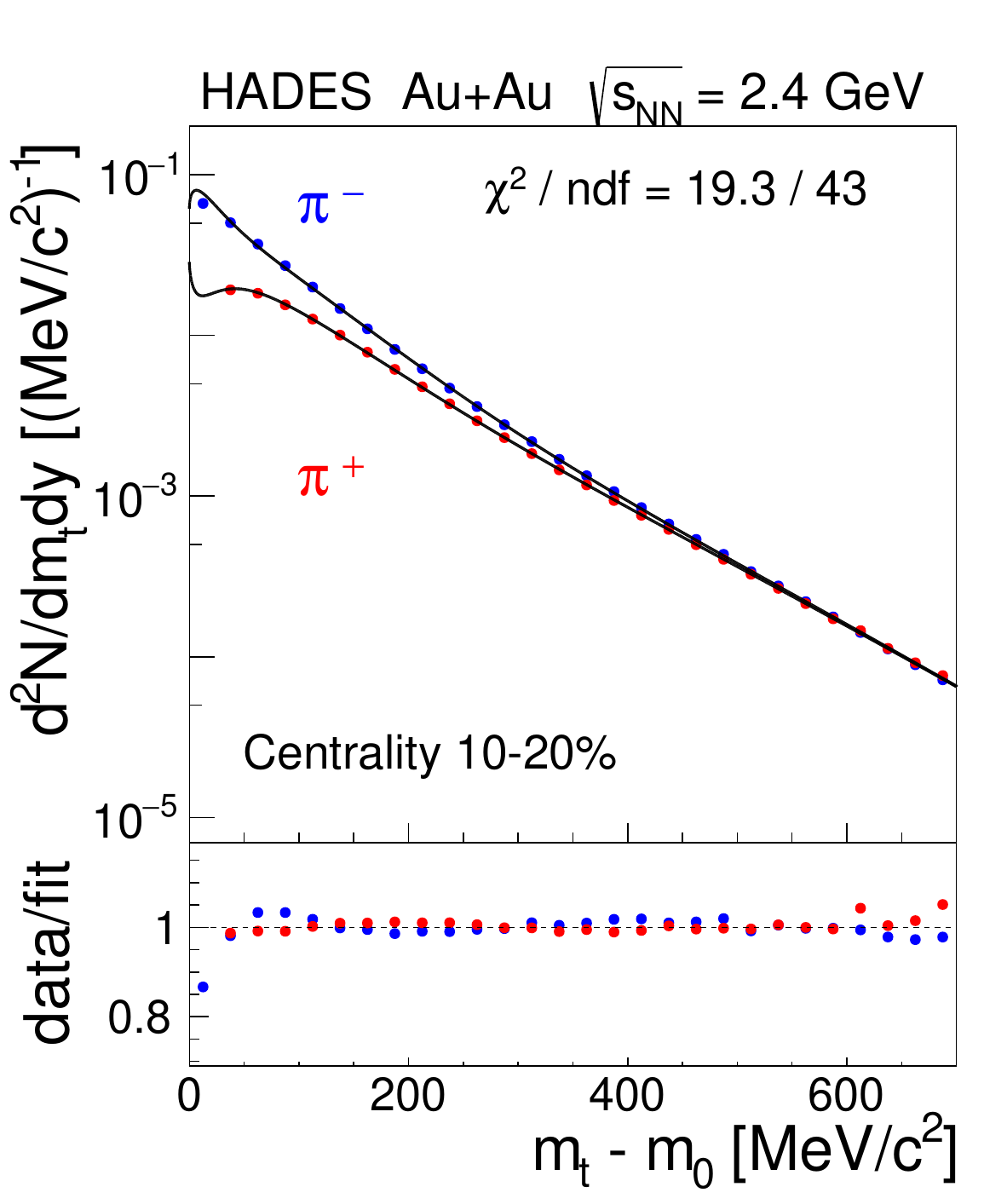}
	\includegraphics[width=0.49\linewidth]{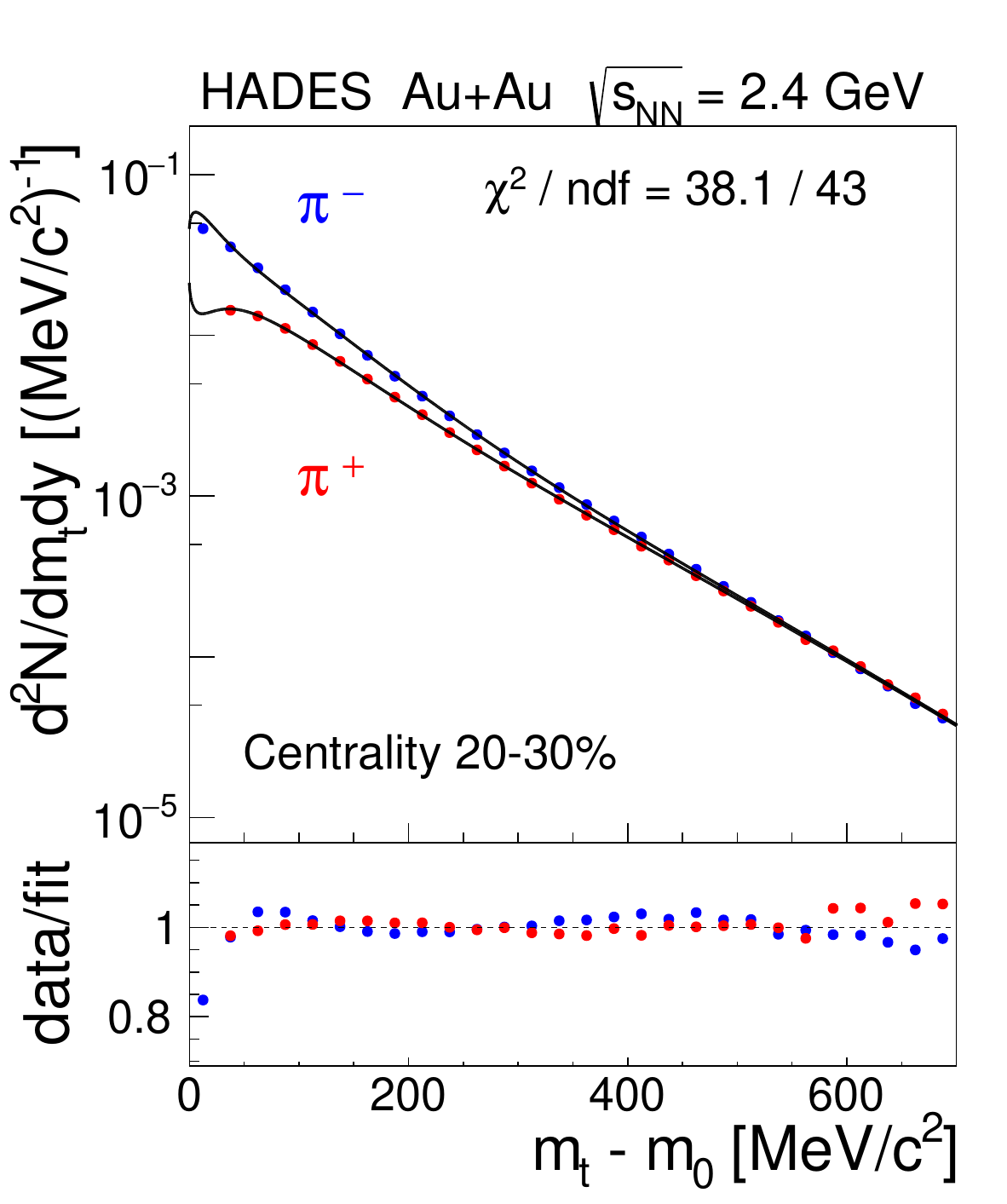}
	\includegraphics[width=0.49\linewidth]{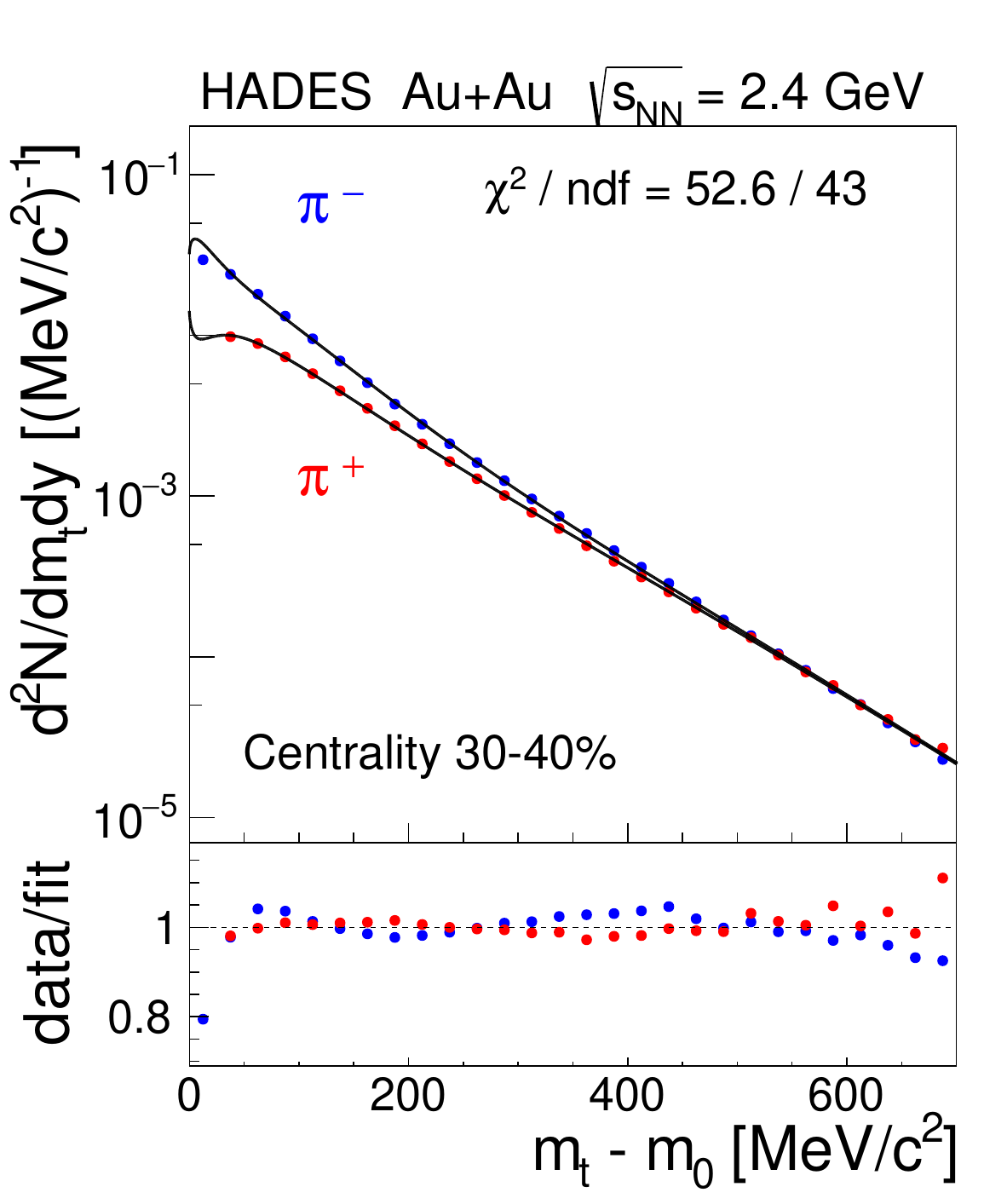}
	\caption{The mid-rapidity transverse-mass spectra of $\pi^-$ (blue points) and $\pi^+$ (red points) measured for the $0-10\%$, $10-20\%$, $20-30\%$,
          and $30-40\%$ centrality classes.  The colored curves are obtained from a simultaneous fit of Eq.~\eqref{eq-fit-mt-Veff} to the data points of both charges.
          The resulting fit parameters and their statistical errors are listed in Tab.~\ref{tab-Vc-HADESdata}.  The lower panels show the ratio of the data divided
          by their corresponding fit function.}
    \label{fig-HADES_allCent}  
\end{figure}	

\begin{table}
 \caption[]{Parameters resulting from fits of Eq.~\eqref{eq-fit-mt-Veff} to HADES mid-rapidity spectra for the $0-10\%$, $10-20\%$, $20-30\%$,
   and $30-40\%$ centrality classes; errors are standard deviations from the fits.  The corresponding mean $A_{\textrm{part}}$ values obtained from a Glauber model fit to the measured hit
   distribution \cite{Adamczewski-Musch:2017sdk} are listed as well.}
 \vspace{0.2cm}
 \centering
	\begin{tabular}{lllll}
		\hline
		Centrality & $\langle A_{\textrm{part}}\rangle$ & $T_{1}$ [MeV]  & $T_{2}$ [MeV]  & $V_{C}$ [MeV] \\ \hline
		$~~0-10\%$ & 303 $\pm$ 12          & 52.5 $\pm$ 0.5 & 97.1 $\pm$ 1.1 & 13.6 $\pm$ 0.6    \\
		$10-20\%$ & 213 $\pm$  12          & 47.9 $\pm$ 0.5 & 89.8 $\pm$ 0.7 & 11.1 $\pm$ 0.7    \\
		$20-30\%$ & 149 $\pm$  10          & 45.9 $\pm$ 0.4 & 85.7 $\pm$ 0.6 & 13.6 $\pm$ 0.6    \\
		$30-40\%$ & 103 $\pm$  ~~8         & 45.1 $\pm$ 0.5 & 82.7 $\pm$ 0.5 &  ~~7.9 $\pm$ 0.8  \\ 
				
		\hline
	\end{tabular}

	\label{tab-Vc-HADESdata}
\end{table}

We have applied the fit procedure presented in the previous section to the charged-pion spectra measured with HADES in Au+Au collisions
at \sqrtsnn\ = 2.4~GeV \cite{Adamczewski-Musch:2020vrg}.   In that experiment, events were selected online with a hardware trigger based on the multiplicity
of hits registered in the HADES time-of-flight detectors.  In total, $10^{8}$ events in the $0-40\%$ most central collisions were selected for the present analysis.
By tracking the charged particles through the HADES magnetic field, their momentum was reconstructed and by recording their time of flight, a velocity was obtained.
Particle identification was finally achieved by cutting on the characteristic momentum-velocity relation.  More details of the event reconstruction,
centrality selection, and pion identification are given in \cite{Adamczewski-Musch:2020vrg}.  The rapidity acceptance of the detector is about 0.65 for charged pions
with slight differences between low-momentum $\pi^{+}$ and $\pi^{-}$ caused by the toroidal geometry of the magnetic field.  To obtain the total pion yields, extrapolation
to zero $m_t$ is required as well as (model-dependent) extrapolation outside of the acceptance.  The uncertainties on the yields are dominated by systematics
of the efficiency correction (3\%) and the extrapolation to full solid angle (5--7\%).  The statistical uncertainties on the yields are negligible
(see Ref.~ \cite{Adamczewski-Musch:2020vrg} for details).

We start by discussing the mid-rapidity charged pion transverse-mass spectra ($|\ycm|< $ 0.05) obtained for the 10\% most central events, where  the centrality of the
Au+Au collisions has been selected by cutting on the number of hits in the HADES time-of-flight detectors.  The spectra for both charges have been fitted simultaneously
in the range of 0.025 $< \mt <$ 0.65~\gevcc\ assuming an attenuated potential $\Veff$ to account for 3D expansion of the fireball (fixing $T_p$ at 142, 125, 115,
and 106~MeV for 0 -- 10\%, 10 -- 20\%, 20 -- 30\%, and 30 -- 40\% centrality classes, respectively).  The data points and the combined fit are displayed
in Fig~\ref{fig-HADES_allCent} as well as the ratios of data to fit functions.  Except for the lowest $m_t$ bin of the negative pion distribution, the ratios
differ by less than 5\% from unity, demonstrating that the fit function based on Eq.~\eqref{eq-fit-mt-Veff} is adequate.  The adjusted parameters $T_1$, $T_2$,
and $V_C$ are listed in Table~\ref{tab-Vc-HADESdata}, together with the average number of participants $\langle A_\textrm{part}\rangle$ obtained from a Glauber
calculation \cite{Adamczewski-Musch:2017sdk}.  In particular, the mean Coulomb potential energy is found to be $V_C = 13.6 \pm 0.6$ MeV.  By integration of the
adjusted functions, the pion yields and their fit errors are obtained, leading to a $\pi^-/\pi^+$ yield ratio of $2.05 \pm 0.10$ at mid-rapidity.

The influence of the Coulomb field on the charged pion spectra has been investigated with a BUU transport model in Ref.~\cite{Teis:1997xi}.   A similar theoretical study,
done with the QMD model, can be found in \cite{UmaMaheswari:1997ig}.  The authors calculated a mean Coulomb potential energy in the range 20 -- 30~MeV for
1\agev\ Au+Au collisions, i.e.\ substantially larger than what we observe in our data.  However, in the BUU calculation, the potential was determined by averaging over
the full space-time of the heavy-ion collision as well as over the full rapidity range, whereas our result is obtained at mid-rapidity.   The fits of the pion spectra
used in the present work are most sensitive to low-momentum pions most likely emitted from $\Delta$ resonances late in the expansion of the fireball and thus emerging
from a dilute charge distribution.  In contrast, the authors of  Ref.~\cite{Teis:1997xi} discuss only pions moving faster than the expanding charge cloud.  Furthermore,
in their model calculation, fast pions are emitted from higher-lying resonances and freeze out earlier, at a density of $\approx 0.8 \rho_0$, seeing thus a much
more compact charge distribution which translates into a correspondingly larger mean Coulomb energy.

\subsection{Centrality dependence of the Coulomb potential} 

\begin{figure*}[htb]
	\centering
	\includegraphics[width=0.30\linewidth]{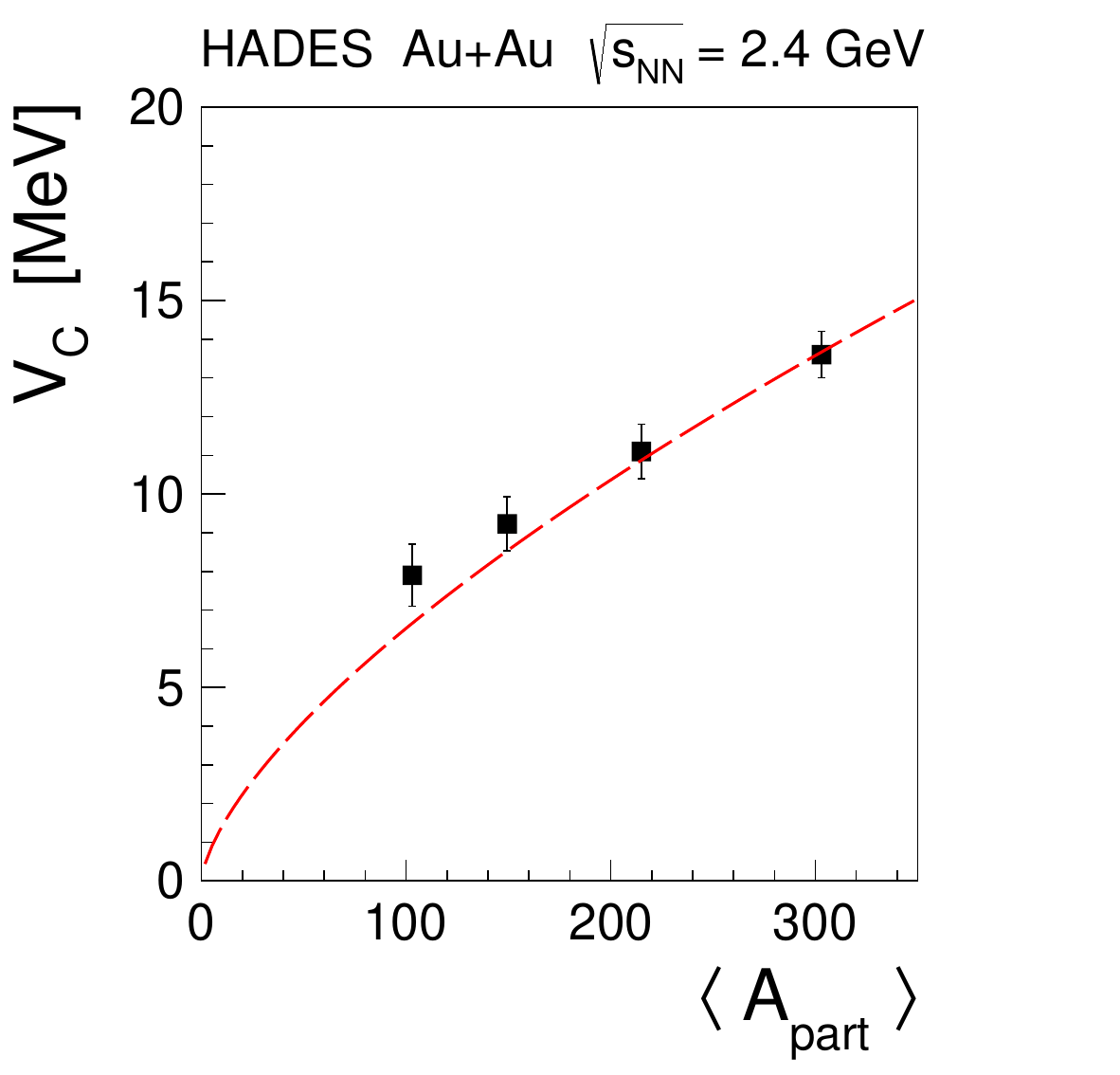}
	\includegraphics[width=0.30\linewidth]{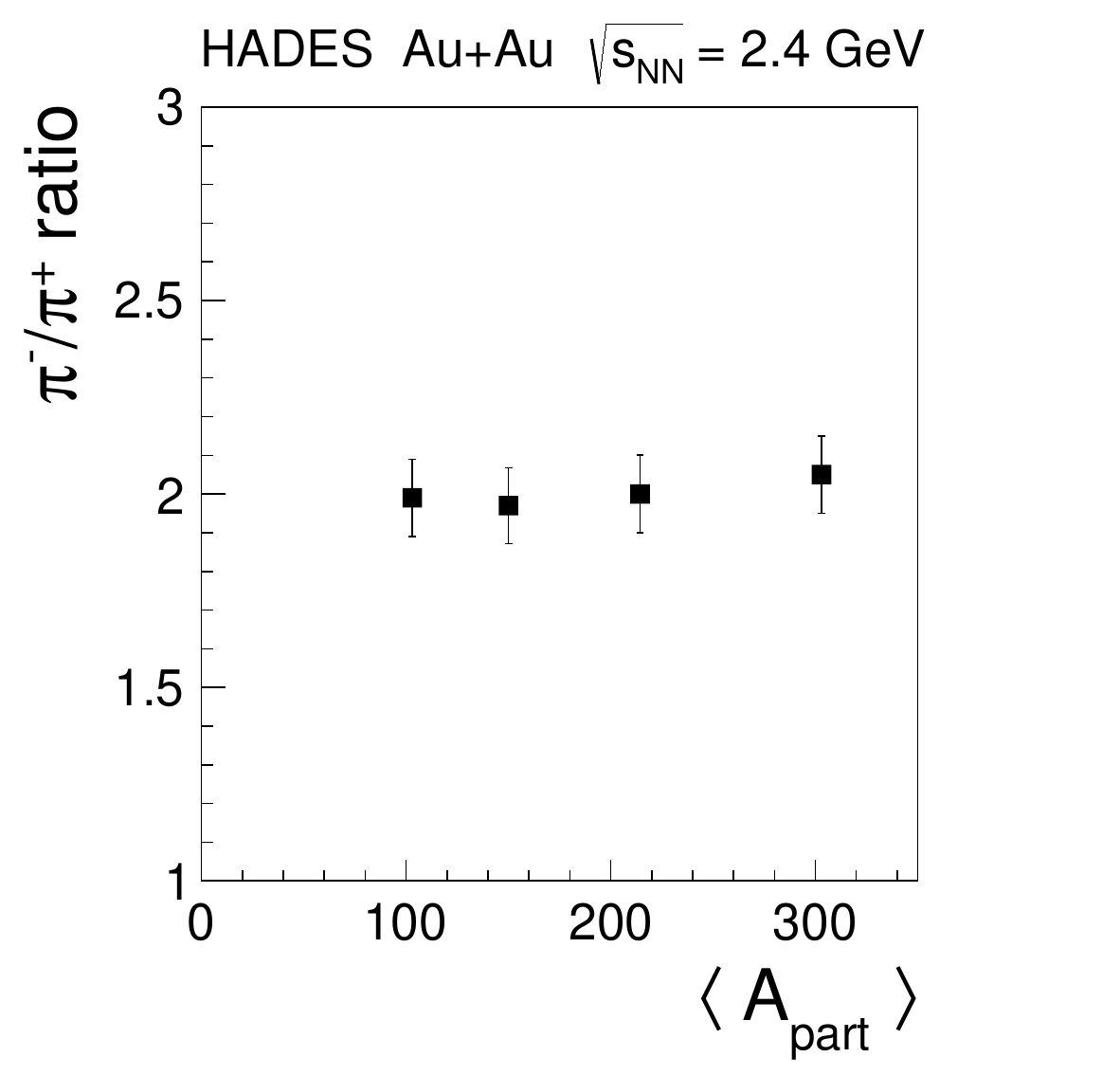}
	\includegraphics[width=0.30\linewidth]{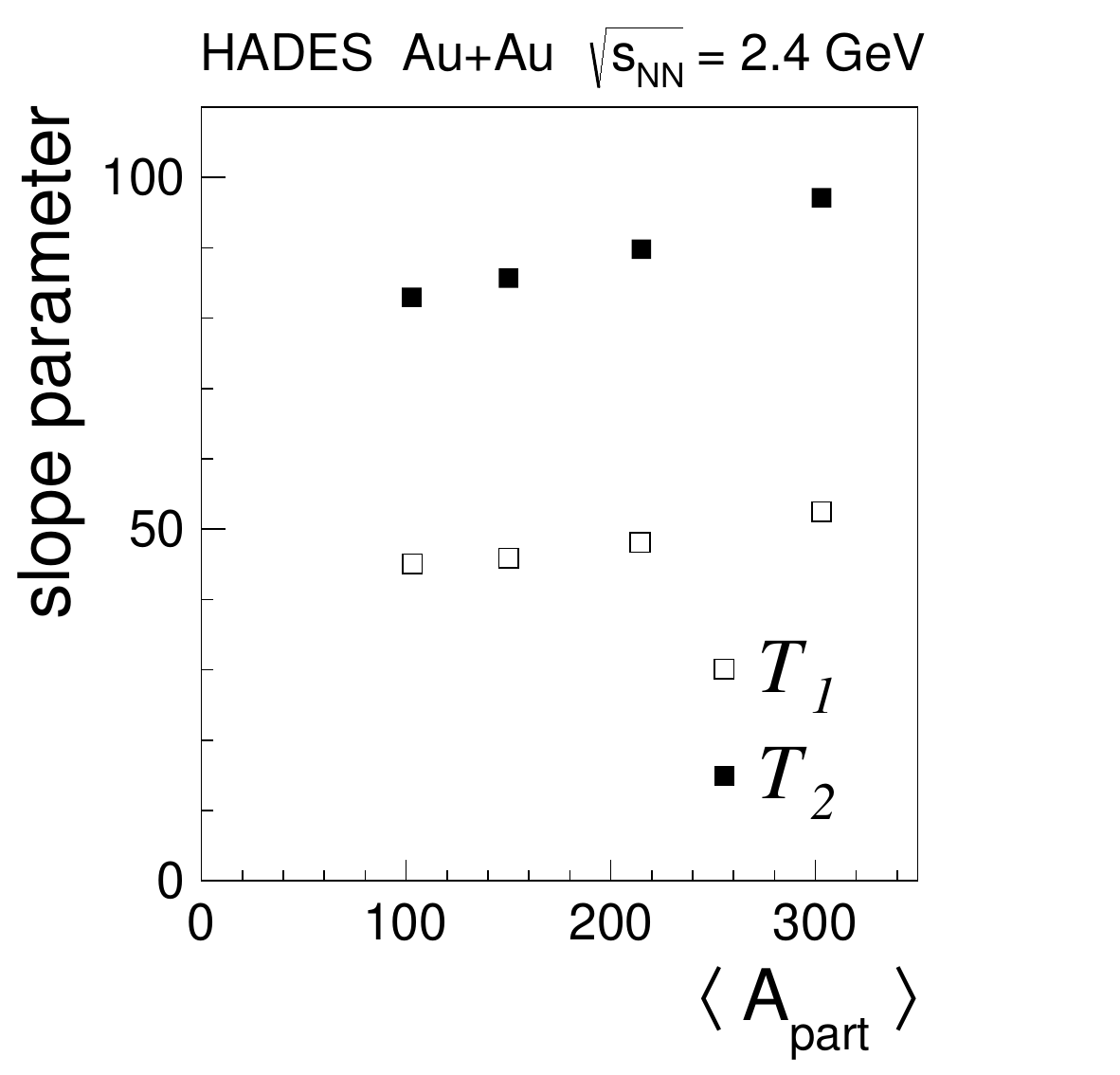}
	\caption{Centrality dependence of various parameters adjusted in the fits to the mid-rapidity pion $m_t$ spectra:  Coulomb potential energy $V_C$ (left),
          mid-rapidity $\pi^{-}/\pi^{+}$ yield ratio (center), and pion inverse slope parameters $T_1$ and $T_2$ (right); vertical bars are fit errors.
          The dashed red line corresponds to a $V_C \propto A_{\mathrm{part}}^{2/3}$ scaling.}
    \label{fig-Hades-fitPar-Apart}  
\end{figure*}

The mid-rapidity charged pion spectra from other centrality classes (10--20\%, 20--30\%, and 30--40\% ) are fitted in a similar fashion.  They are shown in
Fig.~\ref{fig-HADES_allCent} together with the corresponding data-over-fit ratios.  The latter demonstrate again that, except for the lowest $m_t-m_0$ bin,
the data is very well described by Eq.~\eqref{eq-fit-mt-Veff}.  The extracted fit parameters are listed in Table~\ref{tab-Vc-HADESdata} and their evolution with
centrality is summarized in Fig.~\ref{fig-Hades-fitPar-Apart} where $V_C$, the $\pi^{-}/\pi^{+}$ ratio, and the inverse slope parameters $T_1$ and $T_2$ are plotted as
a function of mean $\langle A_{\mathrm{part}} \rangle$.  One observes that the Coulomb potential energy $V_C$ decreases smoothly from central to peripheral collisions.
This decline goes in hand with the decreasing overlap volume, resulting in a smaller fireball and less net charge contributing to the Coulomb effect.  
As can be seen in the figure, the $A_{\mathrm{part}}^{2/3}$ scaling of $V_C$ is not fully realized, suggesting that spectator contributions are indeed present
in the most peripheral events.  Note, however, that the observed effect appears to be weaker than in the schematic model calculations \cite{Gyulassy:1980xb}
underlying Fig.~\ref{fig-Simul-Vc-Apart}.  In Fig.~\ref{fig-Hades-fitPar-Apart}, one furthermore sees that both inverse slope parameters increase with
increasing centrality, whereas the pion ratio remains basically constant.  To avoid complications due to spectator effects, we focus our further analysis
on central collisions. 

\subsection{Pion rapidity densities and total yields}

\begin{figure*}[htb]
	\centering
\includegraphics[width=0.30\linewidth]{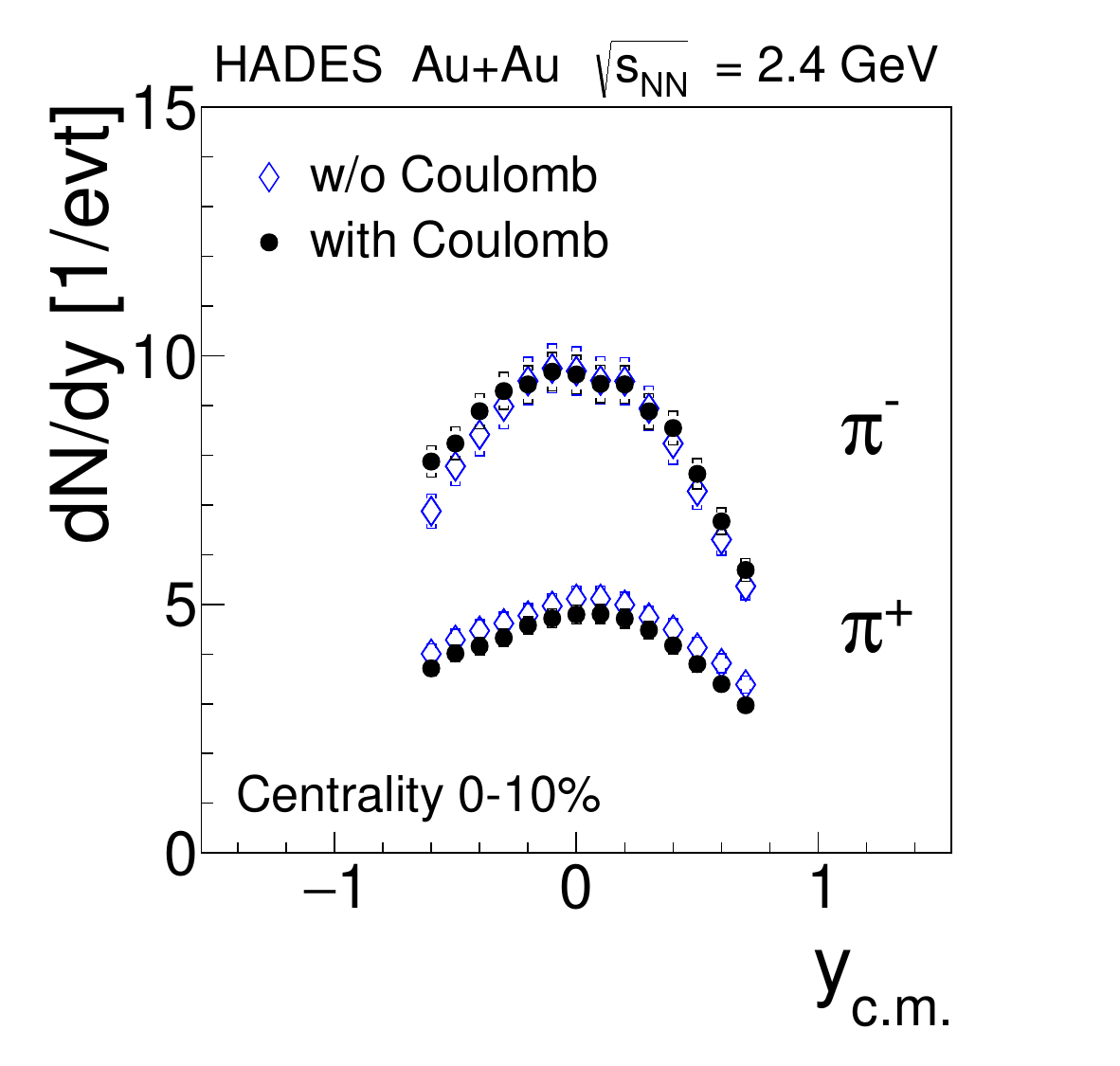}
	\includegraphics[width=0.30\linewidth]{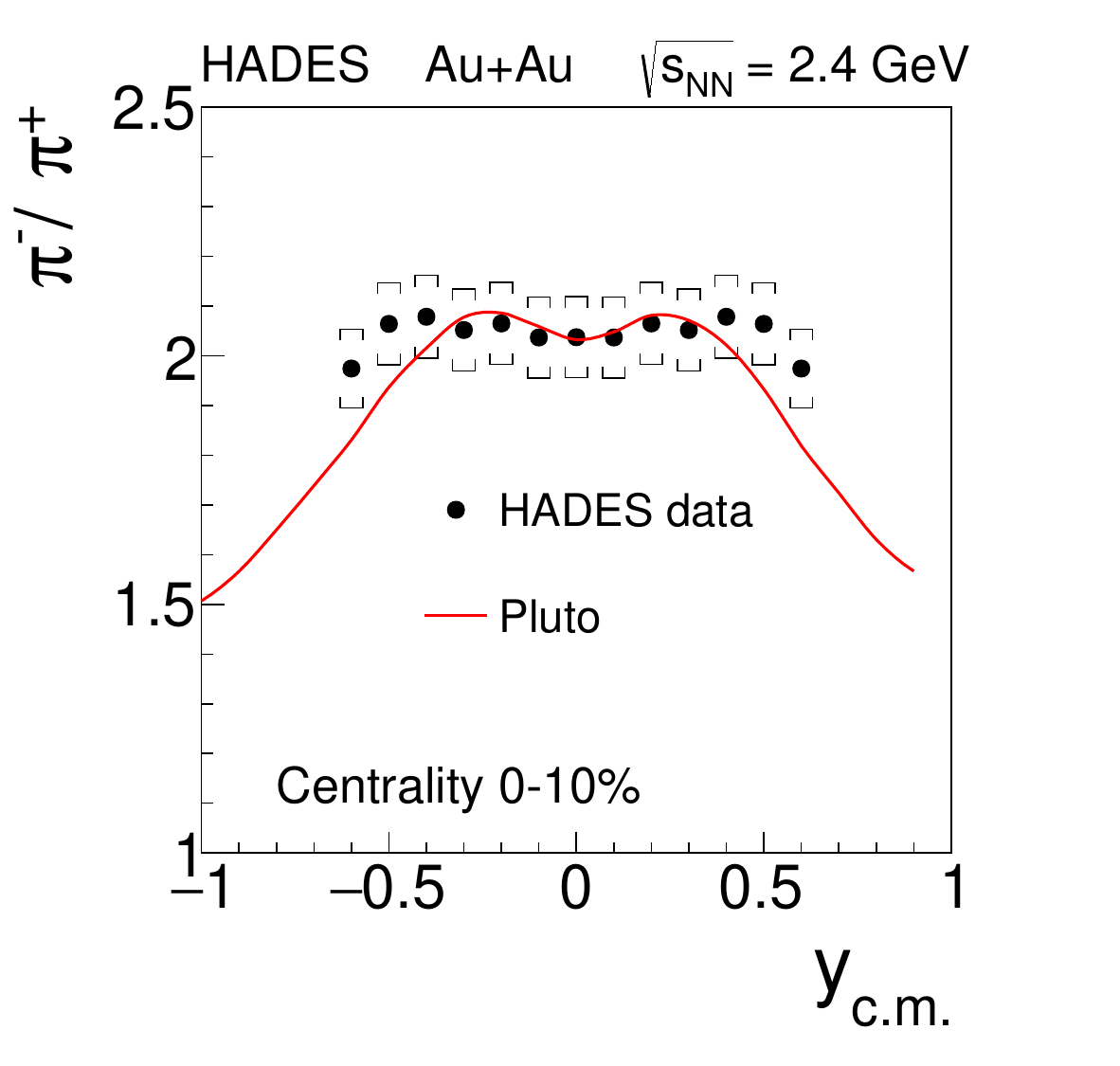}
	\includegraphics[width=0.30\linewidth]{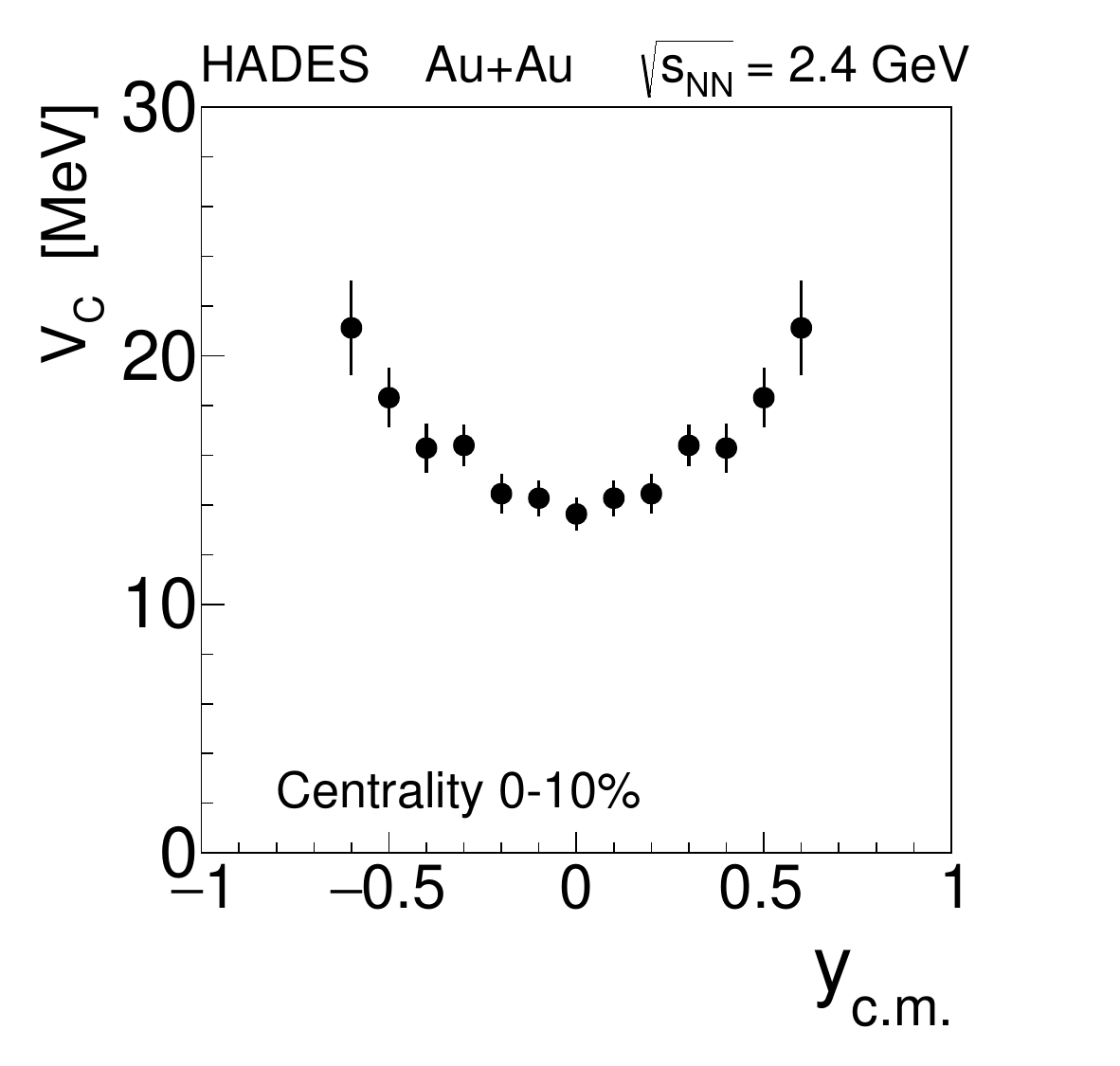}
	\caption{Left: rapidity dependence of charged pion yields obtained by fitting and extrapolating the
          $m_t$ distributions with a Coulomb-modified function (full symbols), respectively a two-slope
          Boltzmann function (open symbols) as used in Ref.~\cite{Adamczewski-Musch:2020vrg}; systematic uncertainties 
          are shown as bars.  Center: symmetrized $\pi^{-}/ \pi^{+}$ ratio for the 0-10\% most central events;
          bars are systematic uncertainties. The red solid curve corresponds to a Pluto simulation of the pion
          ratio done by setting all pion source parameters to the values obtained from our fit to the
          measured $m_t$ spectrum.  Right: extracted and symmetrized Coulomb potential energy $\Vc$.}
    \label{fig-pim2pip_ycm}  
\end{figure*}

The Coulomb field affects not only the transverse momentum but also the rapidity density distributions.  However, here the situation is more complicated due to the observed forward-backward peaked polar anisotropies of the emitted pions \cite{Adamczewski-Musch:2020vrg}.  
As discussed, the extrapolation and integration of the pion $m_t$ spectra with the use of the Coulomb-modified fit function, the $dN/dy$ distributions of $\pi^{+}$ and $\pi^{-}$ can be obtained inside the rapidity range covered by HADES.  The left panel of Fig.~\ref{fig-pim2pip_ycm} shows the integrated pion rapidity density of central events as a function of the center-of-mass rapidity.  
For comparison, the yields based on the Boltzmann fits done in a previous analysis \cite{Adamczewski-Musch:2020vrg} are shown as well. 
It is apparent that the inclusion of the Coulomb field affects the pion yields, decreasing $\pi^{+}$ and increasing $\pi^{-}$.
To obtain the total pion multiplicities per event, an extrapolation of the data points outside of the HADES rapidity coverage is required.  Total pion multiplicities are hence obtained  by extrapolating the measured yields with the help of various transport calculations (see Ref.~\cite{Adamczewski-Musch:2020vrg}), resulting in $M(\pi^{-}) = 17.5 \pm 1.0$ and $M(\pi^{+}) = 8.6 \pm 0.5$, superseding the $17.1 \pm 1.2$ and $9.3 \pm 0.7$ values obtained in Ref.~\cite{Adamczewski-Musch:2020vrg} without considering the impact of Coulomb effects.  The error bars given are systematic and are in both cases dominated by the extrapolation to full solid angle (for details see Ref.~\cite{Adamczewski-Musch:2020vrg}).   
From these numbers, the ratio of the total pion multiplicities is found to be $2.03 \pm 0.14$.  This ratio can be compared to
the predictions of the isobar model \cite{Stock1986}, namely 1.95 when pion production is mediated solely by $\Delta$ resonance excitation, and 1.70 for production solely via $N^{*}$ resonances.  The observed ratio favors clearly a scenario where the $\Delta$ dominates pion production, also in agreement with the findings of a detailed study of resonance excitation and decay in the Au+Au collision system \cite{Adamczewski-Musch:2020edy}.

The rapidity dependence of the measured pion yield ratio, symmetrized around mid-rapidity, is displayed in the middle panel of Fig.~\ref{fig-pim2pip_ycm}, together with a Pluto simulation using source parameters set to the values obtained from the $m_t$ fit, and the polar angular anisotropies $A_2(\pi^{-})$ = 0.66 and $A_2(\pi^{+})$ = 0.54 obtained in Ref.~\cite{Adamczewski-Musch:2020vrg}.
The main features of the observed ratio, in particular the shape of the marked dip around mid-rapidity caused by an interplay of the pion intrinsic angular distribution and the Coulomb field are reproduced. 

The right panel of Fig.~\ref{fig-pim2pip_ycm} shows the dependence of the fitted Coulomb potential energy on the center-of-mass rapidity, symmetrized around mid-rapidity.  It is apparent that $V_C$ has a minimum at mid-rapidity, with the strong rise at backward and forward rapidities being caused by the presence of the target and beam spectator charges, respectively.

\subsection{Geometry of the charge distribution}


The Coulomb potential energy obtained by fitting Eq.~\eqref{eq-fit-mt-Veff} to the $m_t$ spectra can be used to estimate the hadronic density of the pion emitting system at freeze-out.  To do this, we have to make assumptions about the geometry of the charge distribution, on the one side, and of the distribution of pion emission points, i.e.\ shape of the pion source, on the other side.  To get a handle on this problem, we assume that the pions are produced in a spherically symmetric fireball.  Furthermore, we consider two extreme cases: (1) a constant charge density and (2) a Gaussian charge density.

Following Ref.~\cite{Barz:1998ce}, the electrostatic potential energy of a uniformly charged sphere of radius $R_{\circ}$ and total charge $Ze$ is obtained as 
\begin{equation}
V_C(r)=
\begin{cases}
\frac{3}{2}\frac{Ze^2}{R_{\circ}}\left[1-\frac{1}{3}\left(\frac{r}{R_{\circ}}\right)^2\right]  & \mathrm{for}\;\;r<R_{\circ}\\
\frac{Ze^2}{r} & \mathrm{for}\;\;r\geq R_{\circ}\\
\end{cases}
\end{equation}

\noindent
For volume emission, i.e. assuming that the pions decouple instantaneously from the fireball, the average potential energy $\langle V_C \rangle$ is obtained by integrating the potential over a constant source distribution
$f(\mathbf{r})\propto\Theta(R_{\circ}-r)$, where $\Theta$ is the Heaviside function, such that
\begin{equation}
\langle V_C\rangle = \frac{\int d^3\mathbf{r}\;\Theta(R_{\circ}-r) \,V_C(r)}{\int d^3\mathbf{r}\;\Theta(R_{\circ}-r)}= 6/5 \; e^2 \; Z / R_{\circ},
\label{eq:VcHardsphere}
\end{equation}
with $Z = Z_{\mathrm{part}}$ being the number of participating protons, $Z_{\mathrm{part}} = A_{\textrm{part}} \times Z_{\textrm{Au}}/A_{\textrm{Au}}$.  For the 10$\%$ most central Au+Au events, $\langle A_{\textrm{part}} \rangle$ = 303
$\pm$ 12 and $\langle Z_{\textrm{part}}\rangle =122 \pm 5$ ~\cite{Adamczewski-Musch:2017sdk}.  Note that, doing so, we neglect the spectator charges, which is a valid approximation for the most central collisions (see Fig.~\ref{fig-Simul-Vc-Apart}
and discussion above).  If, however, only surface emission is assumed, we must replace in Eq.~\eqref{eq:VcHardsphere} the Heaviside function with the Dirac delta $f(\mathbf{r})\propto\delta(r-R_{\circ})$, recovering the
relation $\langle V_C \rangle = e^2 \, Z_{\textrm{part}} / R_{\circ}$, as used e.g.~in Ref.~\cite{Wagner:1997sa}.    Thus, in the 10\% most central collisions, from the fitted potential $\langle V_C\rangle = 13.6~\pm~0.6$~ MeV,
we find a source radius $R_{\circ} = 15.4 \pm 0.8 ~\textrm{fm}$ ($12.9 \pm 0.7$~fm) for volume (surface) emission, respectively. 

In two-particle momentum correlation analyses, usually a different picture of the fireball at freeze-out is assumed, namely a 3D Gaussian density profile.
The widths, which in the most general case can be different along the three spatial directions, are denoted by $R_{\mathrm{out}}$, $R_{\mathrm{long}}$ and $R_{\mathrm{side}}$.
Restricting ourselves to a spherically symmetric shape, the Coulomb potential energy of a pion emitted at a distance $r$ from the center of a Gaussian charge distribution of width $\sigma$ is given by \cite{Barz:1996gr}
\begin{equation}
\Vc(r) = \frac{Ze^{2}}{r} \; \erf\left(\frac{r}{\sqrt{2} \sigma}\right).
\label{eq:VcGauss0}
\end{equation}

\noindent
Assuming that the pions are emitted from a source following the same density
profile, their average potential energy is given by
\begin{equation}
\begin{split}
\langle \Vc \rangle & = \int d^3 \mathbf{r} \;\;\; \frac{Ze^2}{r} \; \erf\left(\frac{r}{\sqrt{2}\sigma}\right) \;\; \frac{1}{(2\pi)^{3/2}\sigma^3} \;e^{-r^2/(2\sigma^2)} \\
& = \frac{2 Z e^2}{\sqrt{2\pi}} \; \frac{1}{\sigma^3} \int_0^{\infty} r \,dr\, \erf\left(\frac{r}{\sqrt{2}\sigma}\right) \; e^{-r^2/(2\sigma^2)} \\
& = \frac{Z e^2}{\sqrt \pi \sigma} \;.
\end{split}
\label{eq:VcGauss}
\end{equation}

\noindent
With this expression, we obtain for the 10\% most central collisions a Gaussian radius of $\sigma = 7.2 \pm 0.4$~fm.  As discussed in the next section, this value is also consistent with the radii obtained from a two-pion correlation analysis using HADES data \cite{Adamczewski-Musch:2018owj,Adamczewski-Musch:2019dlf}.

By equating the mean potentials obtained by both relations, \eqref{eq:VcHardsphere} and \eqref{eq:VcGauss}, we can define an equivalent hard-sphere radius \cite{GGLP1960} of the Gaussian distribution as
\begin{equation}
  R_{\circ}^{\textrm{eq}} = \frac{6}{5} \, \sqrt{\pi} \, \sigma \simeq 2.13 \; \sigma \;.
\end{equation}

\noindent
In order to make a meaningful comparison between sizes of the fireball calculated in the hard-sphere and Gaussian cases, one can use instead the root mean square
radii $R_{\textrm{rms}}$ of the two density distributions.  In the (volume emission) hard-sphere case one has

\begin{equation}
R_{\textrm{rms}} = \sqrt{\frac{3}{5}} \; R_{\circ} = 12.0 \pm 0.6 \; \textrm{fm} 
\end{equation}

\noindent
and in the Gaussian case

\begin{equation}
R_{\textrm{rms}} = \sqrt 3 \sigma = 12.4 \pm 0.7 \; \textrm{fm} \,.
\label{eq:rmsGauss}
\end{equation}

\noindent

Consequently, expressing $R_{\circ}$ in Eq.~\eqref{eq:VcHardsphere} and $\sigma$ in Eq.~\eqref{eq:VcGauss} in terms of $R_{\textrm{rms}}$, we find that the two density profiles lead to very similar (within $\leq 7$\%) r.m.s.~radii for a given fitted potential $V_C$.  It shows that the Coulomb effect is not very sensitive to details of the charge and pion source distributions, which is due to the long-range nature of the Coulomb force.

\subsection{Coulomb potential from HBT radii}

The Coulomb field of the expanding fireball also acts on the relative momentum of like-sign pion pairs, affecting in a characteristic way the $R_{\pi^{-}\pi^{-}}$ and $R_{\pi^{+}\pi^{+}}$ radii extracted from HBT intensity-interferometry \cite{Baym:1996wk,Barz:1996gr,Barz:1998ce,Baym:1997ce,Hardtke:1997cy}, namely 
\begin{align}
  \frac{R_{\pi^{\pm}\pi^{\pm}}}{R_{\Tilde{\pi}^0\Tilde{\pi}^0}} \approx
\frac{q_\mathrm{i}}{q_\mathrm{f}} = \frac{|{ p}_\mathrm{i}|}{|{p}_\mathrm{f}|} = \sqrt{ 1 \mp 2
\frac{V_{\mathrm{eff}}}{|{p}_\mathrm{f}|} \sqrt{1 +
\frac{m_{\pi}^2}{{p}_\mathrm{f}^2}} + \frac{V^2_{\mathrm{eff}}}{{p}_\mathrm{f}^2}},
\label{eq-hbt-Rratio}
\end{align}
where $q_\mathrm{i}$ ($p_\mathrm{i}$) is the initial relative (absolute) momentum of the pair, $q_\mathrm{f}$ ($p_\mathrm{f}$) is its final, i.e.\ with Coulomb push, relative (absolute) momentum, and $V_{\mathrm{eff}}$ is the effective Coulomb potential energy. Relative and absolute pair momenta are formed from the momenta of the two individual pions, $p_\mathrm{1,(i,f)}$ and $p_\mathrm{2,(i,f)}$ as $p_\mathrm{i,f} = (p_\mathrm{1,(i,f)} + p_\mathrm{2,(i,f)})/2$ and $q_\mathrm{i,f} = (p_\mathrm{1,(i,f)} - p_\mathrm{2,(i,f)})/2$.

Typically, one has
$V_{\mathrm{eff}} / k_\mathrm{t} \ll 1$, where $k_t$ is the average transverse momentum of the two pions forming the pair, so that the second-order term $V_\mathrm{eff}^2/p_\mathrm{f}^2$ is very small.
\footnote{In our analysis, the smallest used average $k_\mathrm{t}$ value is about 80 Mev$/c$, leading to $V_{\mathrm{eff}} / k_\mathrm{t} < 1/5$. Furthermore, one has  $p_\mathrm{f} = \sqrt{k_\mathrm{t}^2 + k_\mathrm{l}^2} > k_\mathrm{t}$.}
Neglecting this term, the source radii of constructed neutral-pion pairs (denoted here by $\Tilde{\pi}^{0} \Tilde{\pi}^{0}$) had been obtained in Refs.~\cite{Adamczewski-Musch:2018owj,Adamczewski-Musch:2019dlf} as the average of the squared charged radii
\begin{equation}
R^{2}_{\Tilde{\pi}^{0} \Tilde{\pi}^{0}} = \frac{1}{2}
(R^{2}_{\pi^{-}\pi^{-}} + R^{2}_{\pi^{+}\pi^{+}}) \,.
\label{R-pi0-tilde-HBT}
\end{equation}
\noindent
Likewise, by taking the difference of the squared radii, the effective potential $\Veff$ can be determined as a function of $m_t$, where $m_t$ is the average pion transverse mass around mid-rapidity 
\begin{equation}
V_\mathrm{eff}(m_{t}) = \frac{m_{t}^2 - m_{\pi}^{2}}{4m_{t}} \;\; \frac{R^{2}_{\pi^{-}\pi^{-}} - R^{2}_{\pi^{+}\pi^{+}}}{R^{2}_{\Tilde{\pi}^{0} \Tilde{\pi}^{0}}} \,.
\label{eq:Veff-HBT}
\end{equation}

\noindent
Fitting the resulting $\Veff(m_t)$ with Eq.~\eqref{eq-Veff-cases}, one obtains from the HBT radius another estimate of the Coulomb potential energy $V_C$.  
The situation is, however, more complex in HBT interferometry because the full 3D analysis of the emitting source gives various radii, namely $R_{\mathrm{long}}$, $R_{\mathrm{side}}$, $R_{\mathrm{out}}$, and, finally, $R_{\mathrm{int}}$ which can be considered as an average over the three axes (see Ref.~\cite{Adamczewski-Musch:2019dlf} for a definition).  Moreover, the interpretation of these radii in terms of a volume of homogeneity rather than a geometric charge volume has to be kept in mind when comparing with the $R_{\textrm{rms}}$ radius of the single-pion analysis.

\begin{figure}[htb]
	\centering
	\includegraphics[width=0.8\linewidth]{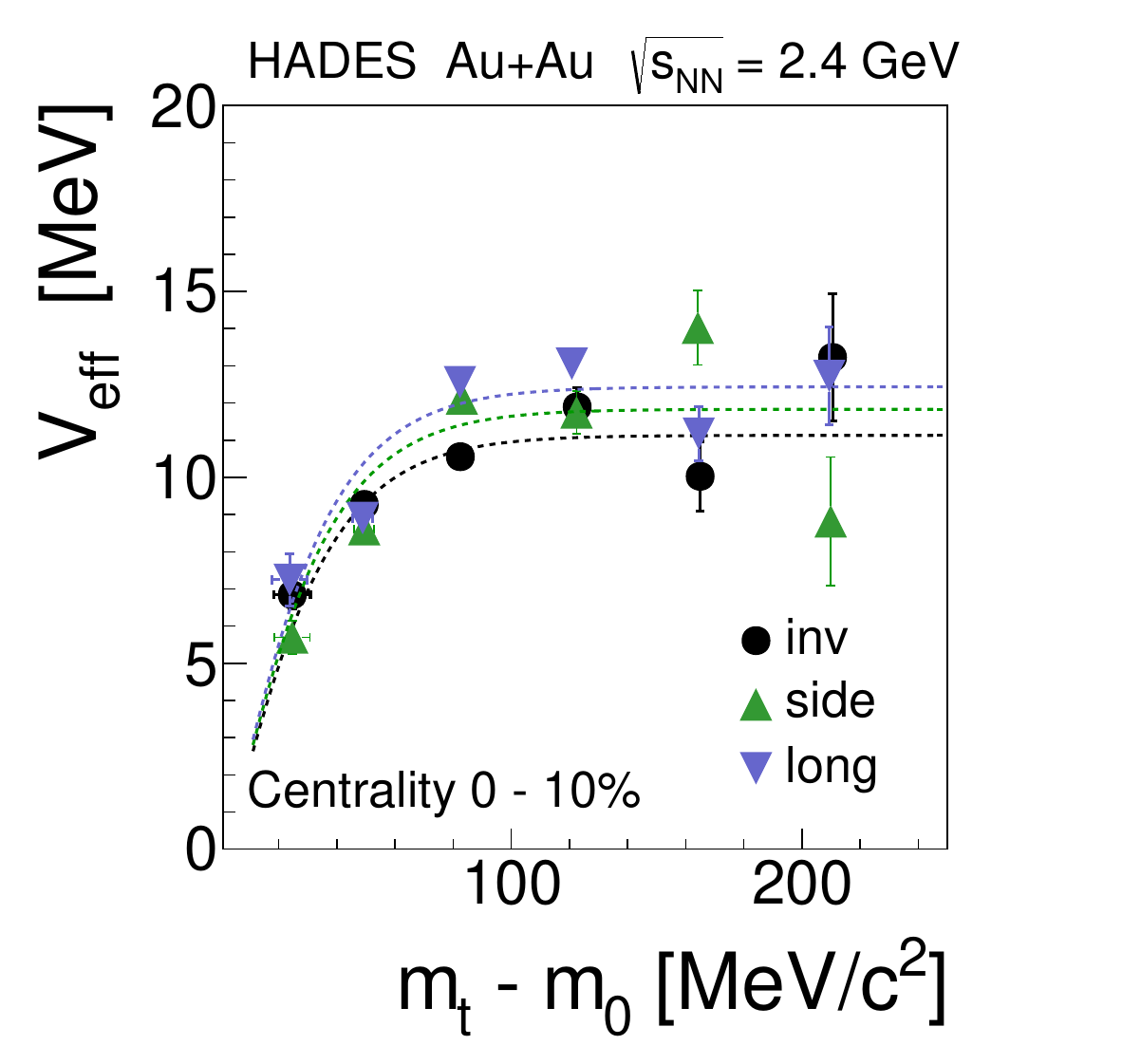}
     	\caption{The effective Coulomb potential energy ($V_{eff}$) obtained from the charged-pion HBT
          analysis plotted as a function of the average pion transverse mass
          ($-0.35 \leq y^{\pi}_{\textrm{c.m.}} \leq 0.35$) in the 0 - 10\% centrality class. Data points correspond
          to different HBT radii (black: $R_\mathrm{inv}$, green: $R_\mathrm{side}$, blue: $R_\mathrm{long}$);
          dashed curves are the fits done with the 3D form of Eq.~\eqref{eq-Veff-cases}.}
    \label{fig-HBT-fitVc}  
\end{figure}

\begin{figure}[htb]
	\centering
	\includegraphics[width=0.8\linewidth]{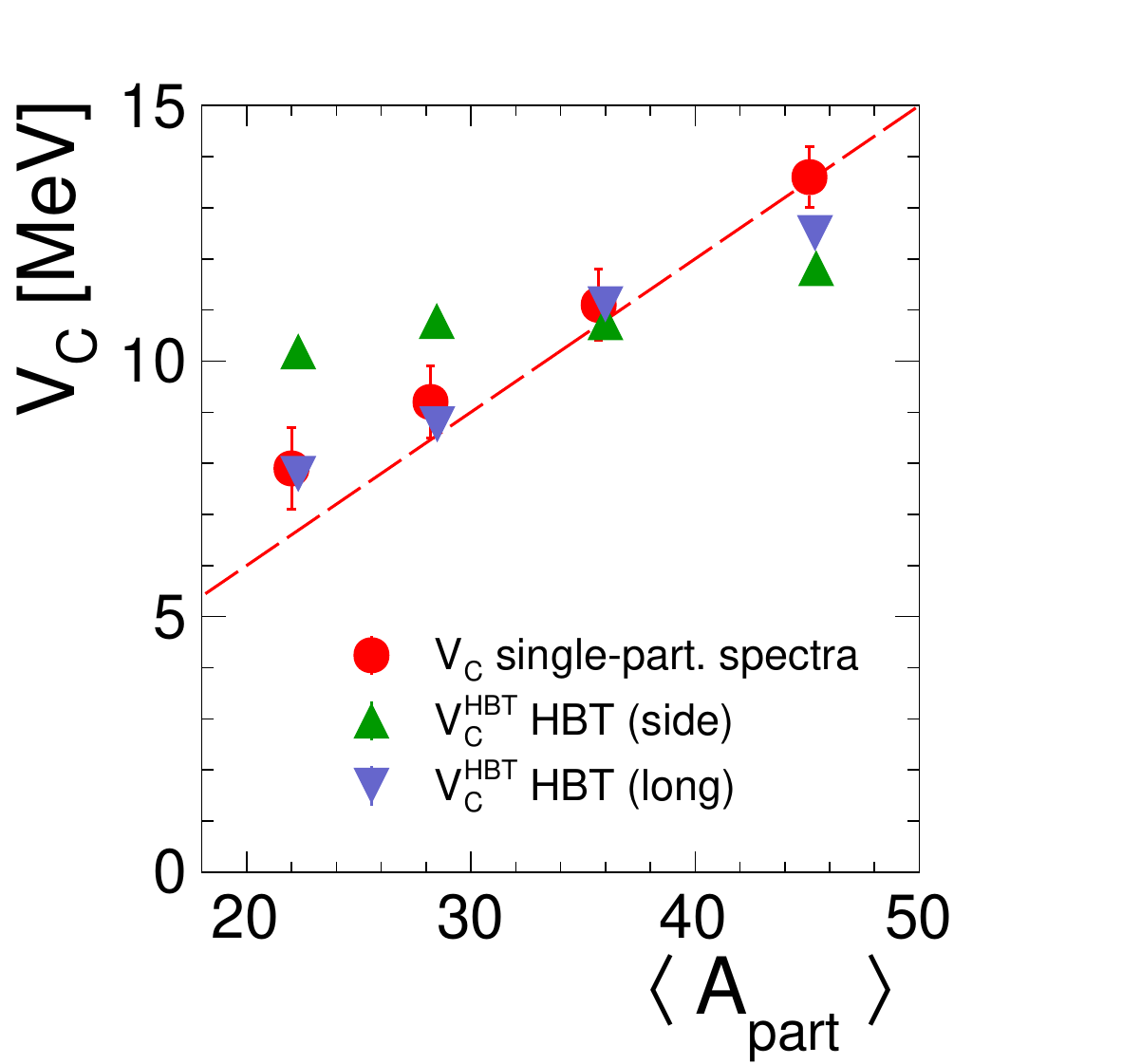}
	\caption{Values of $V_\mathrm{C}$ extracted from the charge-sign difference of the HBT radii using
          Eq.~\eqref{eq:Veff-HBT} ($V^{HBT}_\mathrm{C}$, triangles) compared to the values obtained from
          the single-particle ($m_\mathrm{t}-m_0$)-spectra (red circles) as a function of $\langle A_\mathrm{part}\rangle^{2/3}$.
	  Only $V^\mathrm{side}_\mathrm{C}$ (green up-pointing triangles) and $V^\mathrm{long}_\mathrm{C}$
          (blue down-pointing triangles) are considered. The dashed line denotes the expected linear
          trend corresponding to $A_{\mathrm{part}}^{2/3}$ scaling, adjusted to hit the most central data point.
	}
    \label{fig-HBT-Vc-vs-Apart}  
\end{figure}

Applying the procedure to the various charged-pion HBT radii published in Ref.~\cite{Adamczewski-Musch:2019dlf}, the $V_\mathrm{eff}^\mathrm{HBT}$ values depicted in Fig.~\ref{fig-HBT-fitVc} were obtained as a function of the mean pion $m_t$.
Results are shown for the 0 -- 10\% centrality, a pion rapidity coverage of $-0.35 \leq \ycm \leq 0.35$, and for $m_\mathrm{t} < 250$ MeV$/c^{2}$ only, as at larger transverse mass the sensitivity of the HBT approach to $V_\mathrm{C}$ dwindles.  All effective potentials, $V^\mathrm{inv}_\mathrm{eff}$, $V^\mathrm{side}_\mathrm{eff}$, and $V^\mathrm{long}_\mathrm{eff}$, were found to range between 5 and 15 MeV.  Just as the various radii are not equal, the corresponding $V_\mathrm{eff}$ also differ.  We have excluded $V^\mathrm{out}_\mathrm{eff}$ from
the discussion because of the more complex influence of the Coulomb field on $R_{\textrm{out}}$ (discussed in Ref.~\cite{Barz:1998ce}) which renders Eq.~\eqref{eq:Veff-HBT}
inapplicable.

To study the centrality dependence of the potential, the fits were performed for all four measured centrality classes.  The values of $V^\mathrm{inv}_\mathrm{C}$, $V^\mathrm{side}_\mathrm{C}$ and $V^\mathrm{long}_\mathrm{C}$ are similar compared to each other with variations of less than 25\% within one centrality class.  Again, we do not show the 'out' direction
and focus here on 'side' and 'long' noting that $V^\mathrm{side}_\mathrm{C}$ varies over all centrality classes by less than 20\% while $V^\mathrm{long}_\mathrm{C}$ increases from peripheral to central collisions by more than 50\%.  Both are plotted in Fig.~\ref{fig-HBT-Vc-vs-Apart} as a function of $\langle A_\mathrm{part}\rangle^{2/3}$ (green and blue triangles) together with the values obtained from the single-pion spectra (shown also in Fig.~\ref{fig-Hades-fitPar-Apart}).
The Coulomb potential energy resulting from both methods, single-particle spectra and HBT, are in reasonable agreement.  In particular, the HBT 'long' direction values are very close to the single-pion spectra result.
The similarity of the Coulomb potential energies obtained from the transverse-mass spectra and from the HBT analysis supports the assumption that the region in which the pions freeze-out and the region of homogeneity of two-pion correlations  overlap.
In Fig.~\ref{fig-HBT-Vc-vs-Apart}, a deviation between the expected $A_\mathrm{part}^{2/3}$ scaling and the measured values of $\Vc$ is again visible towards the more peripheral
collisions.  In the figure this scaling is shown by the dashed red line normalized to the most central data point.  We can only speculate that this behavior may be caused by the
contribution of the spectator protons to the Coulomb field, as was already pointed out when discussing Fig.~\ref{fig-Simul-Vc-Apart}.

\subsection{Net-baryon density and chemical potential}

\begin{figure*}[htb]

    \centering
	\includegraphics[width=0.30\linewidth]{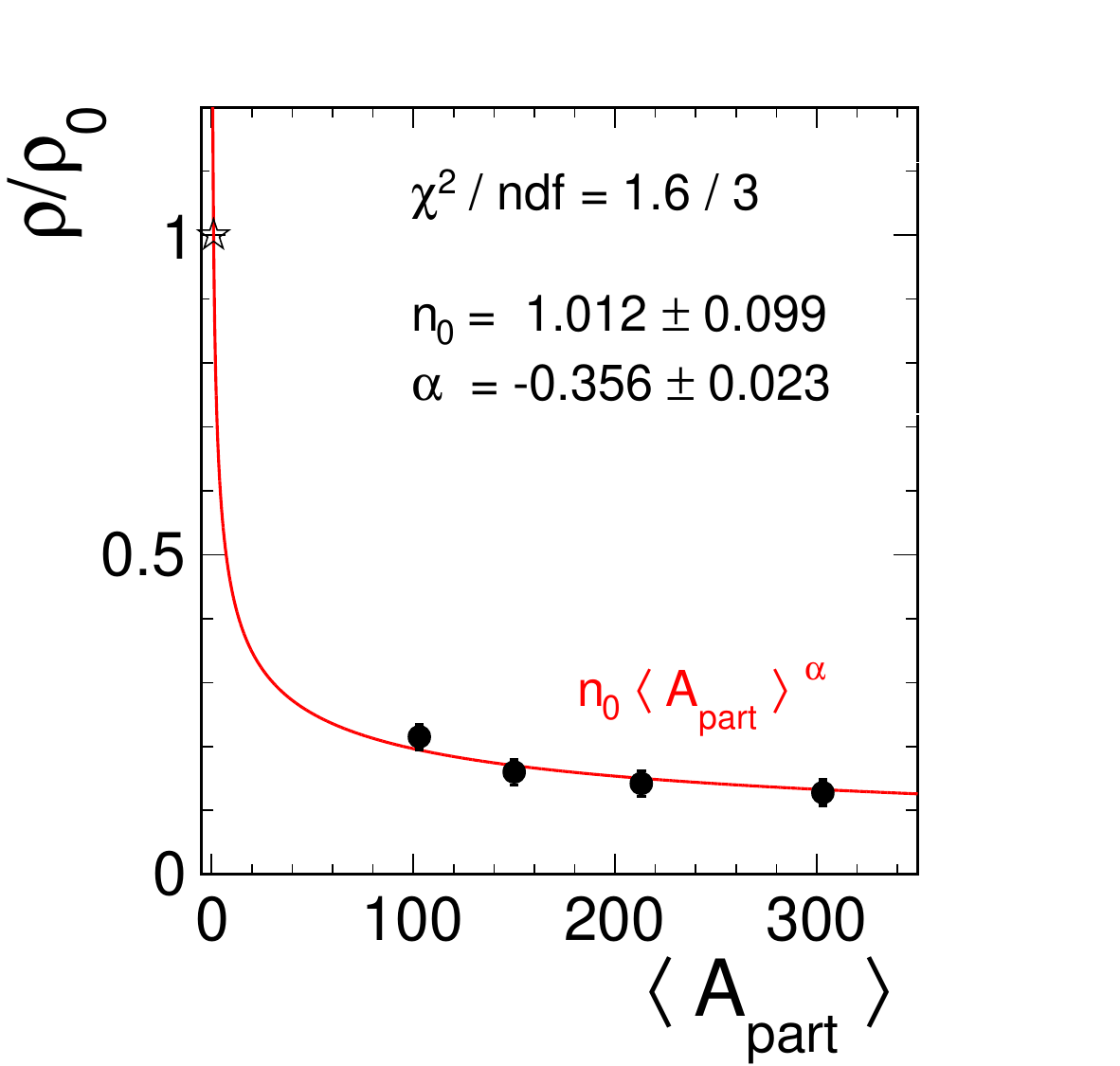}
	\includegraphics[width=0.50\linewidth]{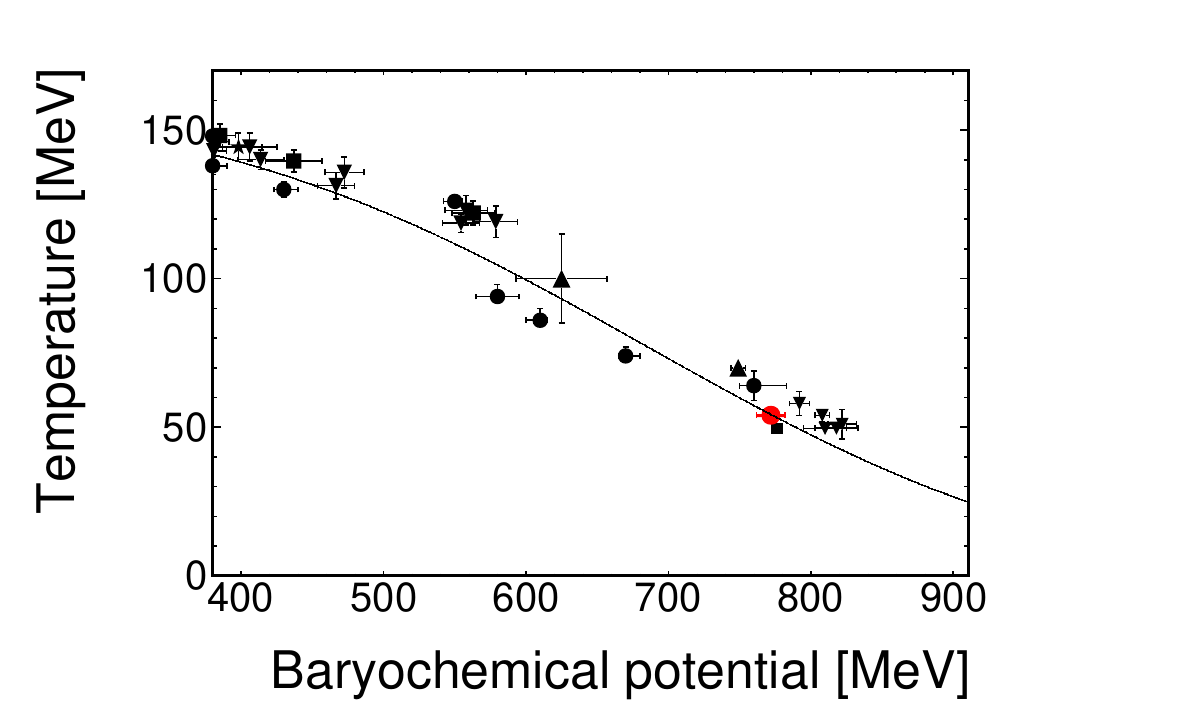}
	\caption{Left: hadron density $n_0 = \rho/\rho_0$ derived from the observed mean potential $V_C$ as a function of $A_{\mathrm{part}}$. The red solid curve corresponds to a power-law fit to the densities; the star symbol indicates nuclear ground-state density.  Right: chemical freeze-out points in the temperature vs. baryochemical potential plane, including the extracted $\mu_B(T)$ curve of Ref.~\cite{Andronic:2017pug}.  The red symbol corresponds to the chemical potential obtained from the Coulomb potential energy fitted for the most central collisions when using  $T_{\textrm{HRG}} = 54.5$~MeV (see text for details). The black symbols stand for freeze-out points extracted in various SHM analyses \cite{Andronic:2017pug,Becattini:2016xct,Cleymans:2005xv,STAR:2017sal,FOPI:2007btf,HADES:2015oef}. }
    \label{fig-phasediagram}  
\end{figure*}

Having confirmed that the source radius extracted from single-particle spectra is in fair agreement with the HBT radii, we use this radius as an approximation of the fireball size and estimate the corresponding baryon density $\rho_B$ at freeze-out.  Setting the baryon number equal to $A_{\mathrm{part}}=303$ and using the hard-sphere radius for volume emission, $R_{\circ} = 15.4\pm0.8$~fm, the average baryon density is found to be $\langle \rho_{B} \rangle = A_{\mathrm{part}}/\frac{4}{3}\pi R_{\circ}^3=0.12 \pm 0.02~\rho_{0}$, where $\rho_{0}$ is the nuclear saturation density.  Alternatively, we can assume a Gaussian source distribution with a maximal density at the center ($r = 0$) of $\rho_{B,max} = 0.32\pm0.05~\rho_0$, corresponding to an average density $\langle \rho_B \rangle = \rho_{B,max}/2^{3/2} = 0.11\pm0.02~\rho_0$, very close to the hard-sphere value.  Note that using a Gaussian source also implies volume emission of the pions.  This is warranted because the Coulomb effects manifest themselves mostly at the very low momenta where pion re-absorption in the medium is known to be weak \cite{Buss:2006,Metag:2017}.
With the $V_C$ values extracted for the four centrality classes, we obtain the evolution of $\rho/\rho_0$ with the number of participating nucleons $A_{\textrm{part}}$ shown in Fig. \ref{fig-phasediagram}(left).  The power law $\rho/\rho_0 = n_0 \, A_{\textrm{part}}^{\alpha}$ fitted to the extracted densities, shown as red solid line, yields $\alpha \simeq -1/3$ and $n_0 \simeq 1$.

The freeze-out density can be converted to the baryon chemical potential $\mu_B$ of an ideal hadron resonance gas (HRG) of temperature $T$ by the following relation (see e.g.~\cite{Landau}):
\begin{equation}
\mu_B = m_N+T\ln{\left(\frac{\rho_B}{g}\left(\frac{m_N T}{2\pi}\right)^{-3/2}(\hbar c)^{3}\right)}\;,
\label{eq:def_mub}
\end{equation}
where, $m_N$ is the nucleon mass and $g=4$ is the degeneracy factor for nucleons.
In order to apply this relation, the temperature $T$ needs to be fixed. However, an unambiguous determination of $T$ from the pion spectra is challenging due to distortions by resonance decays and collective expansion (see Ref.~\cite{Harabasz:2020sei} for a recent study).  Therefore, we calculate $\mu_B$ with the temperature extracted from a systematic study of freeze-out points \cite{Andronic:2017pug} where $T$ and $\mu_B$ have been parameterized as a function of the reaction center-of-mass energy.  The resulting relation is displayed in Fig.~\ref{fig-phasediagram}(right), together with freeze-out points extracted in various SHM analyses of measured hadron yields in heavy-ion collisions over a range of
collision energies spanning from 2~\gev\ up to 2.76~\tev\ \cite{Andronic:2017pug,Becattini:2016xct,Cleymans:2005xv,STAR:2017sal,FOPI:2007btf,HADES:2015oef}. The temperature corresponding to our collision energy is found to be $T_{\textrm{HRG}} = 54.5 \pm 0.5$~MeV and, inserted into Eq.~\eqref{eq:def_mub}, it leads to a baryochemical potential of $\mu_B = 772 \pm 10$~MeV.  The resulting freeze-out point is indicated by the red filled symbol in the phase diagram.  
Note finally that the source radius $R_{\circ} = 15.4\pm 0.8$ found in our analysis is also very close to the freeze-out radius $R_{\textrm{f.o.}}$ typically extracted from SHM analyses
of the measured hadron yields, e.g.~Ref.~\cite{Harabasz:2020sei} found $R_{\textrm{f.o.}}$ = 16 fm.

\subsection{Collision energy dependence of the Coulomb effect}

\begin{figure}[htb]
	\centering
	\includegraphics[width=0.70\linewidth]{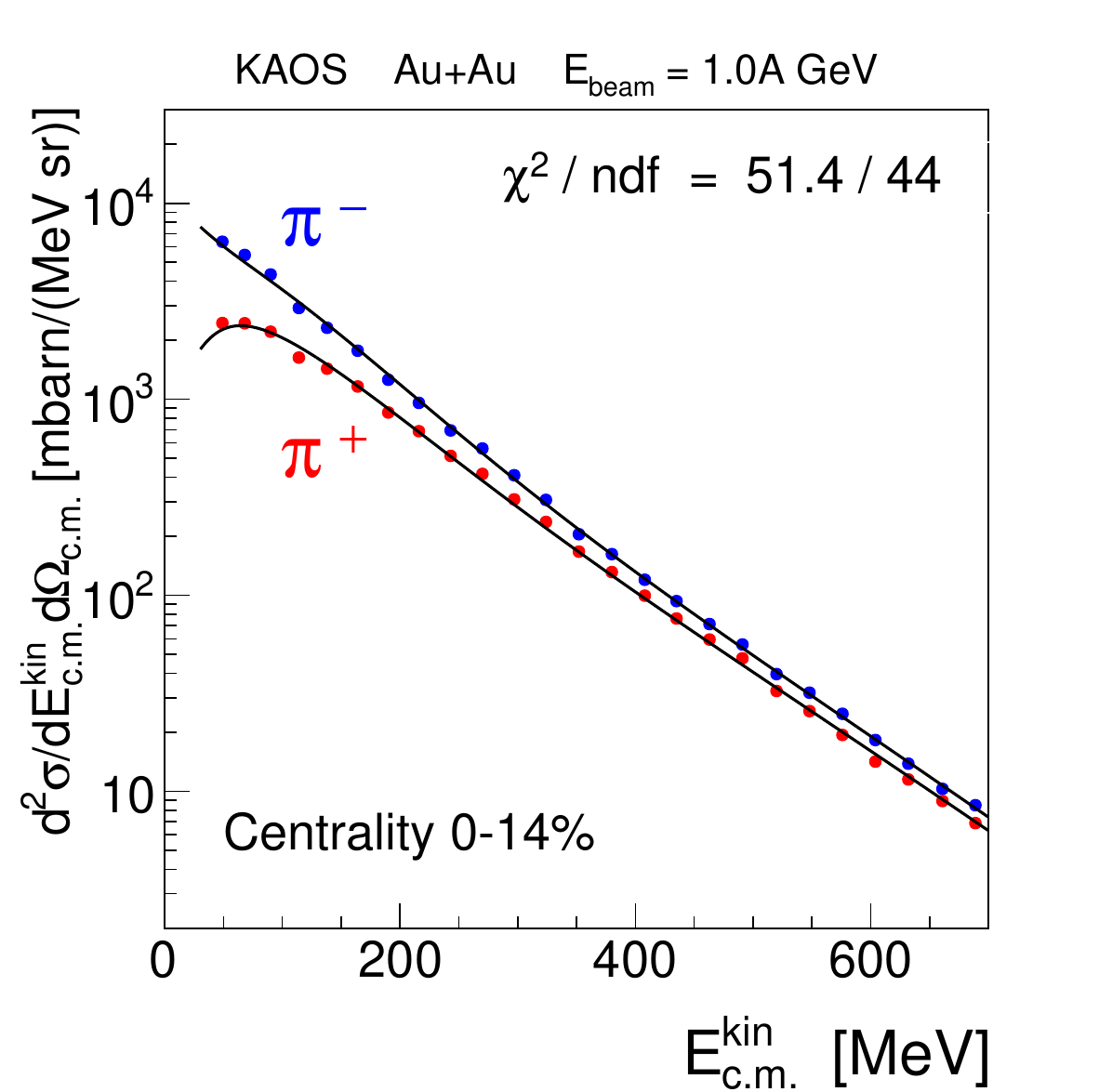}
	\caption{Coulomb-modified two-slope Boltzmann fits to charged-pion kinetic energy spectra measured by the KaoS
          experiment \cite{Wagner:1997sa} in the 14\% most central 1\agev\ Au+Au collisions,
          setting $T_p = 120$~MeV.  These data have been measured at a laboratory polar angle of
          $\theta_{\mathrm{lab}} = 44^{\circ} \pm 4^{\circ}$, corresponding to $\ycm \simeq 0.15$.
	}
    \label{fig-KAOS}  
\end{figure}	

\begin{figure}[htb]
	\centering
	\includegraphics[width=0.49\linewidth]{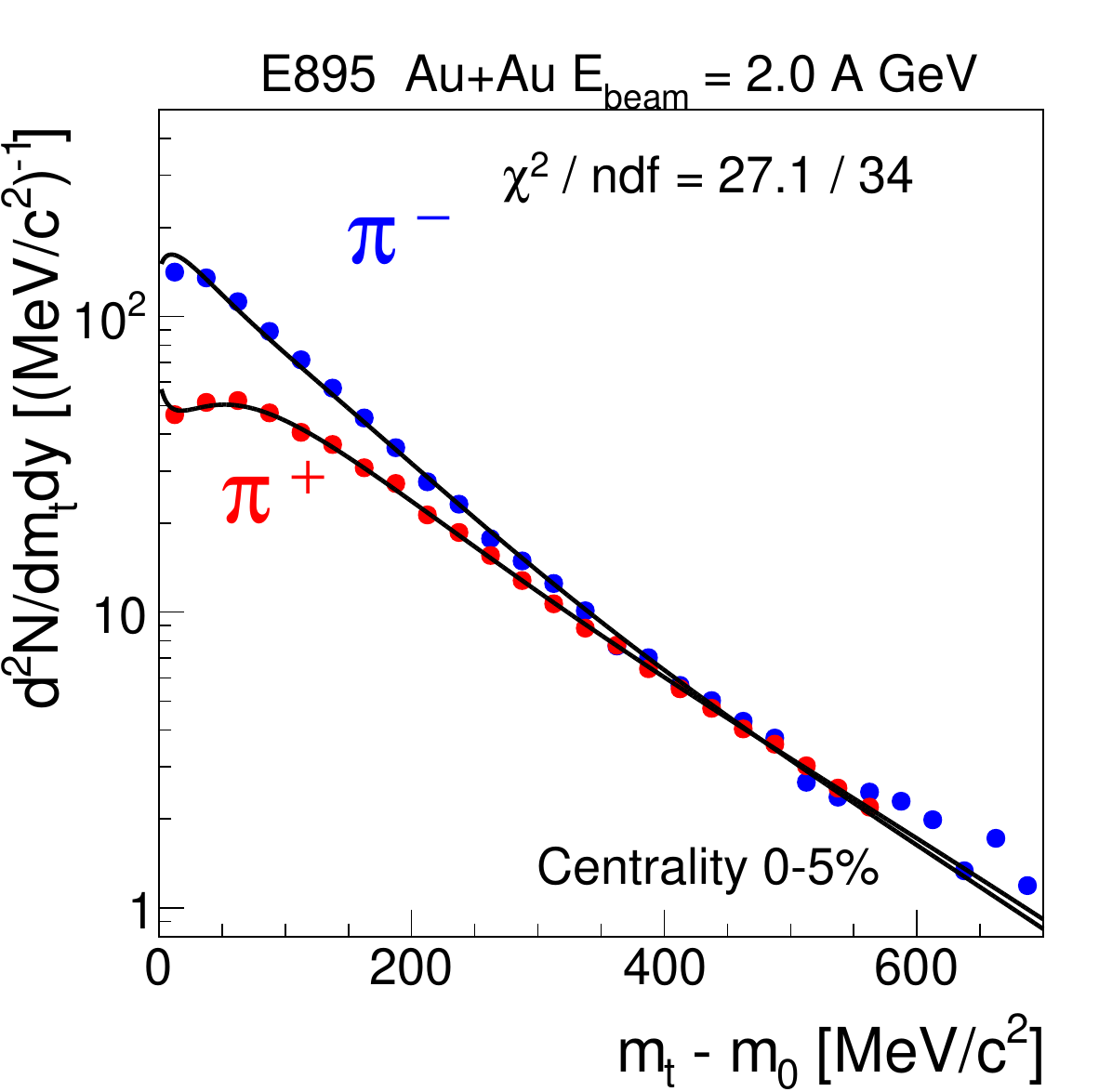}
	\includegraphics[width=0.49\linewidth]{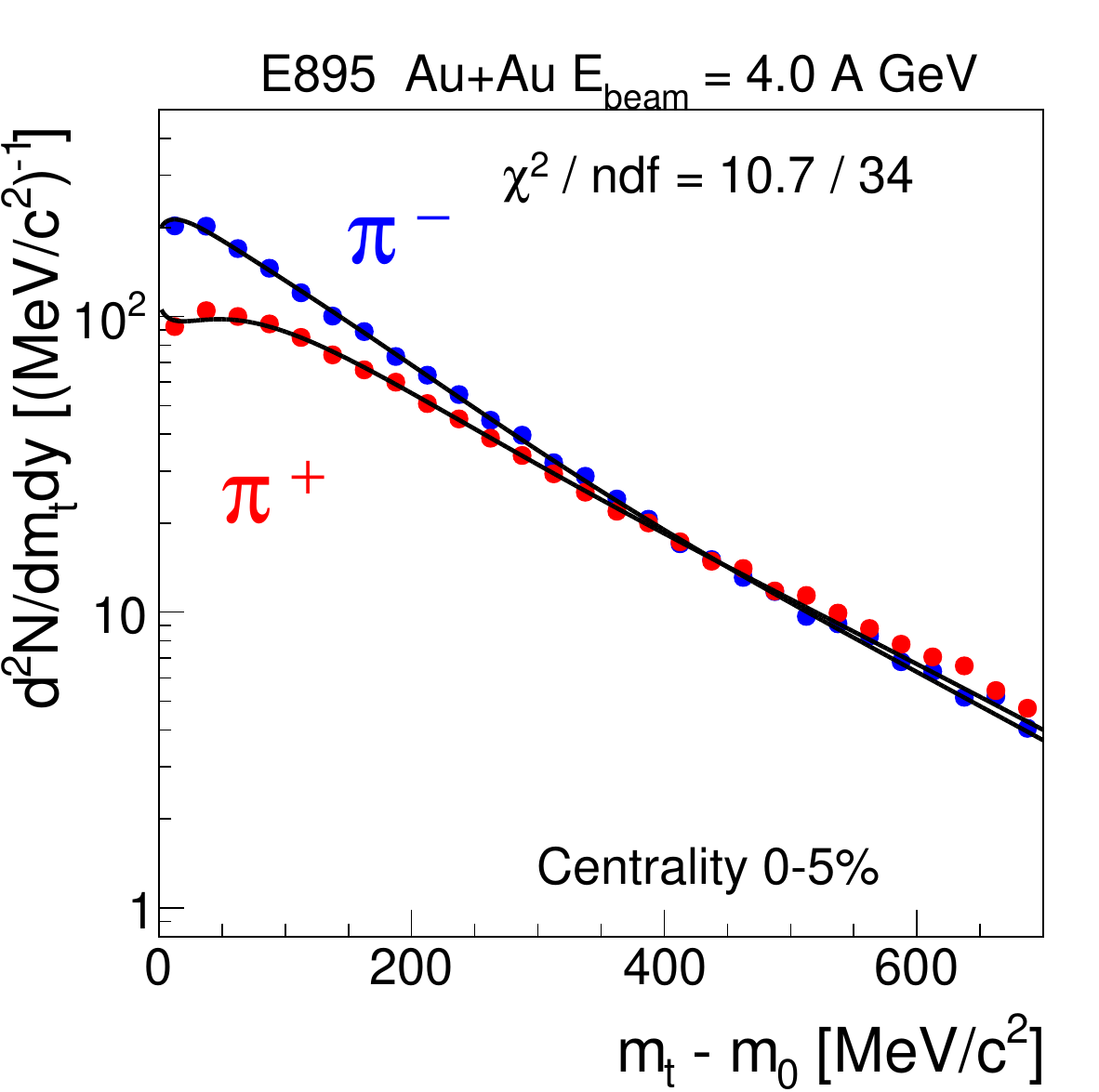}
	\includegraphics[width=0.49\linewidth]{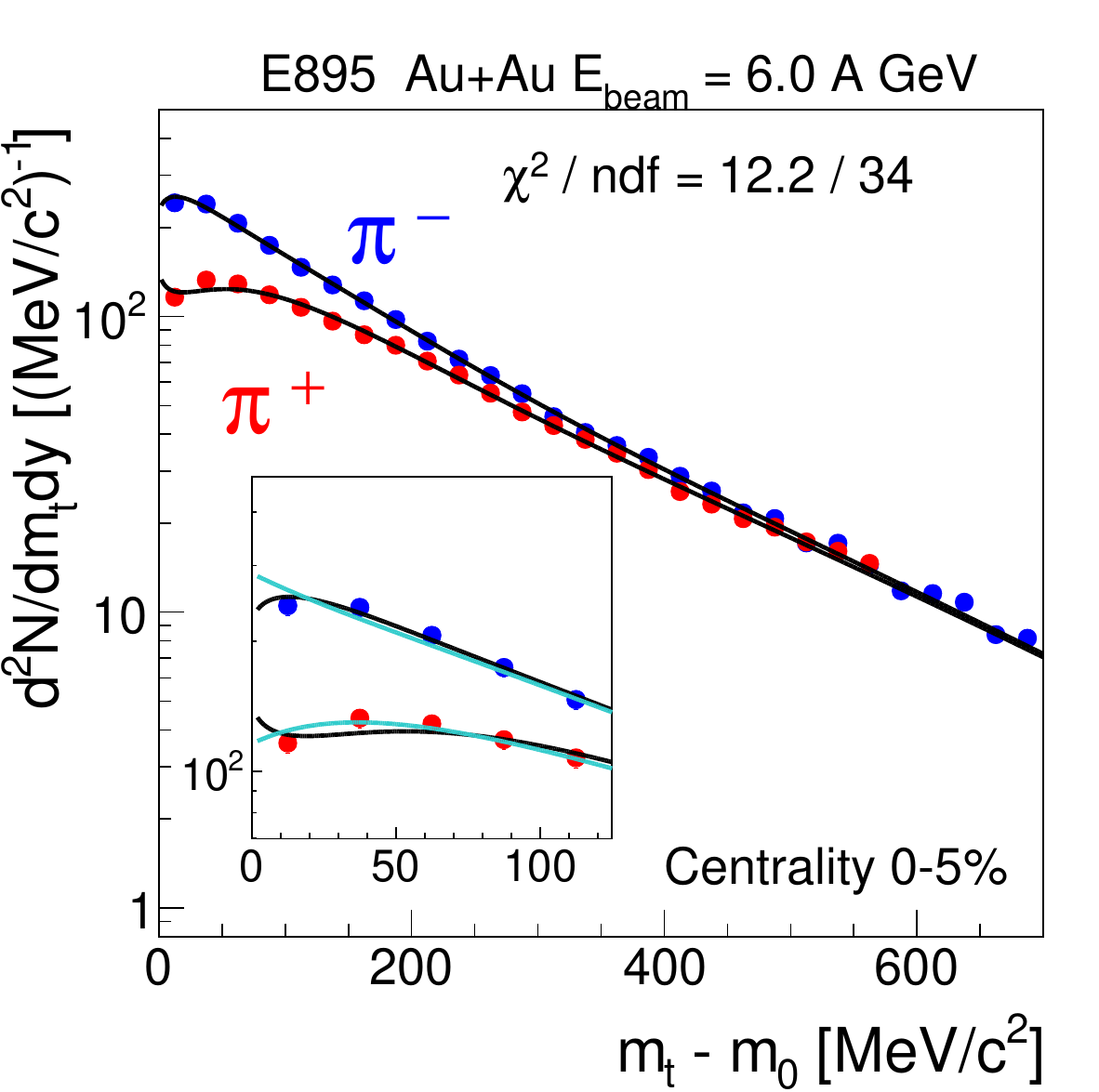}
	\includegraphics[width=0.49\linewidth]{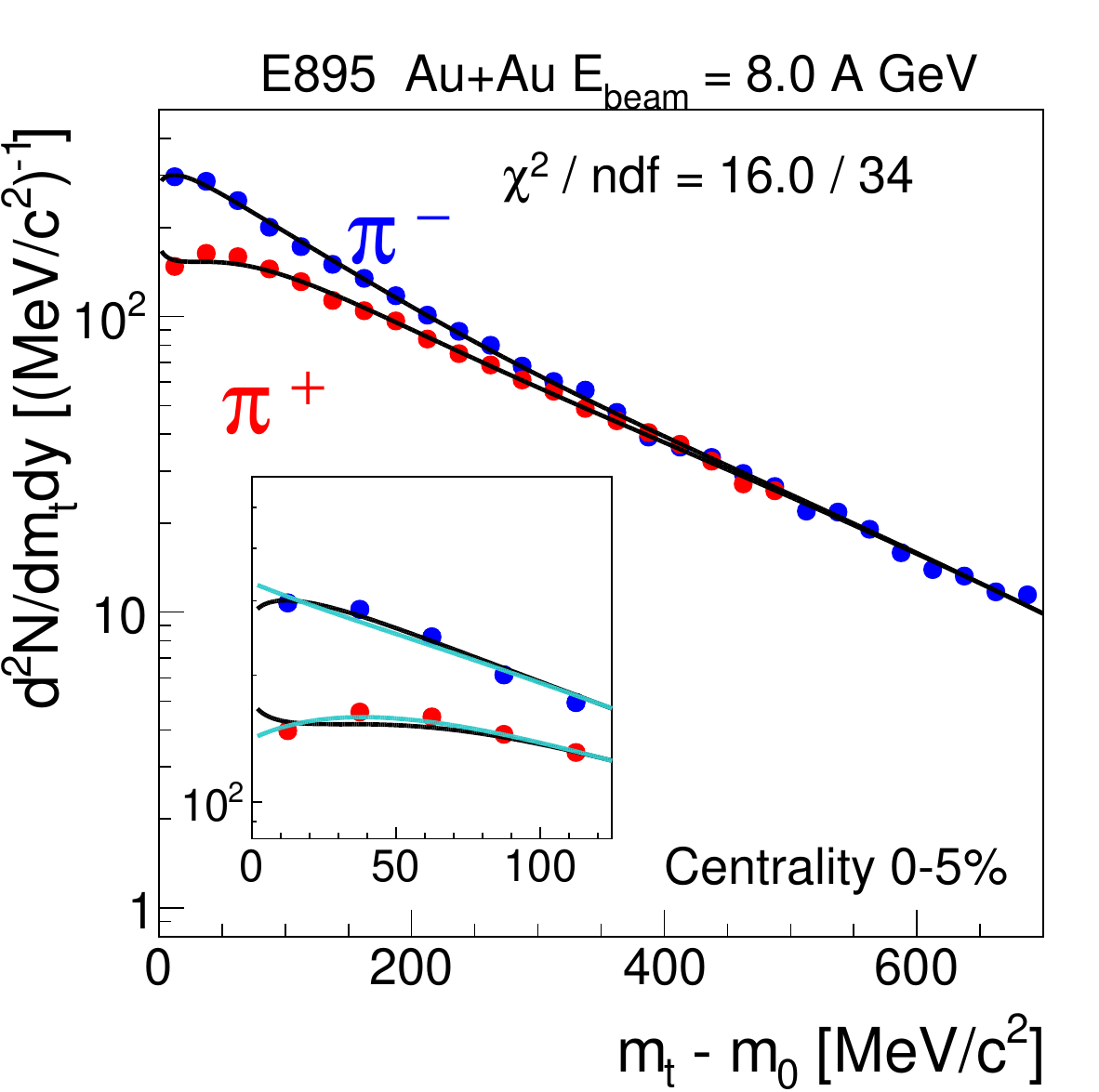}
	\includegraphics[width=0.49\linewidth]{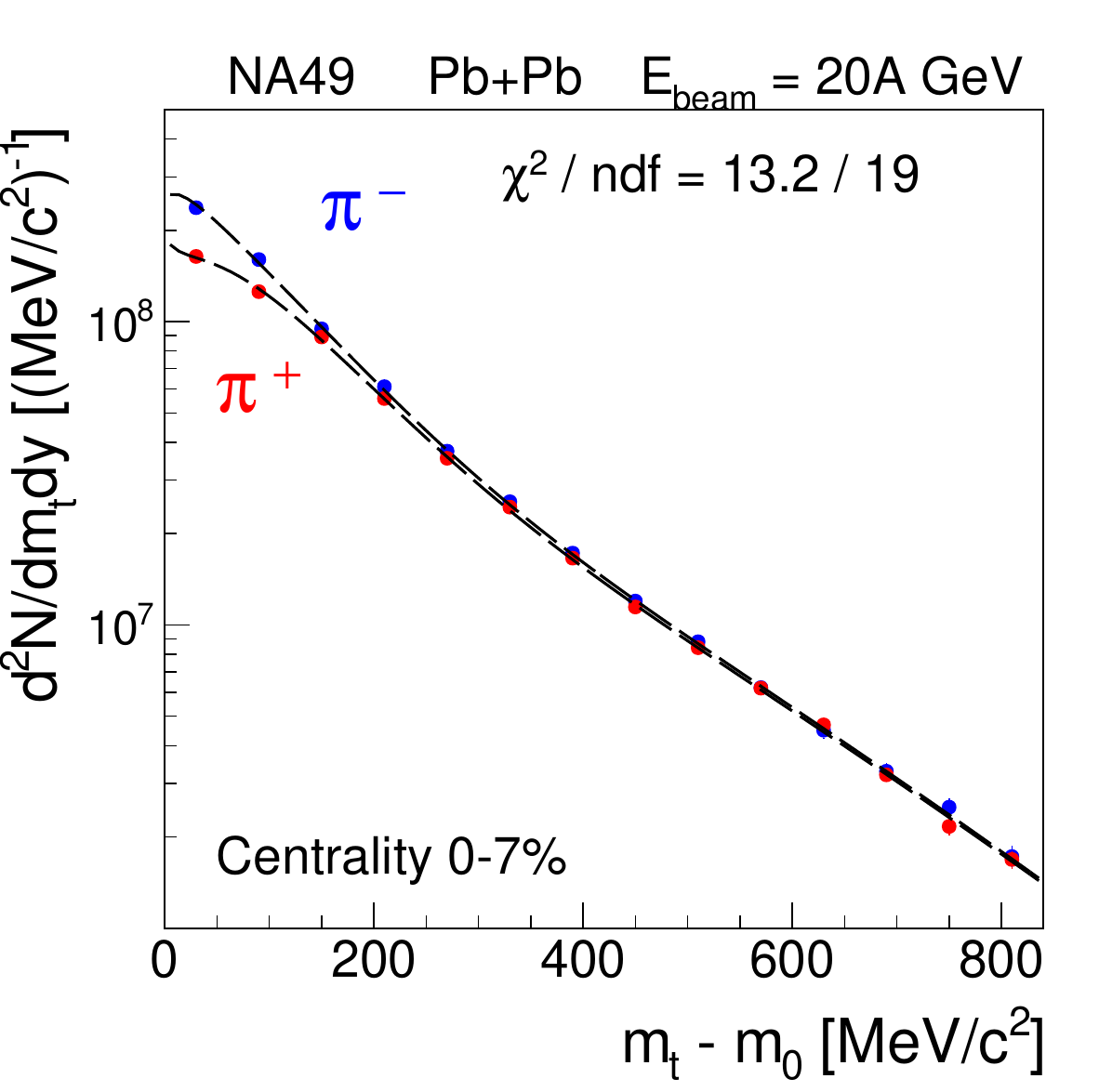}
	\includegraphics[width=0.49\linewidth]{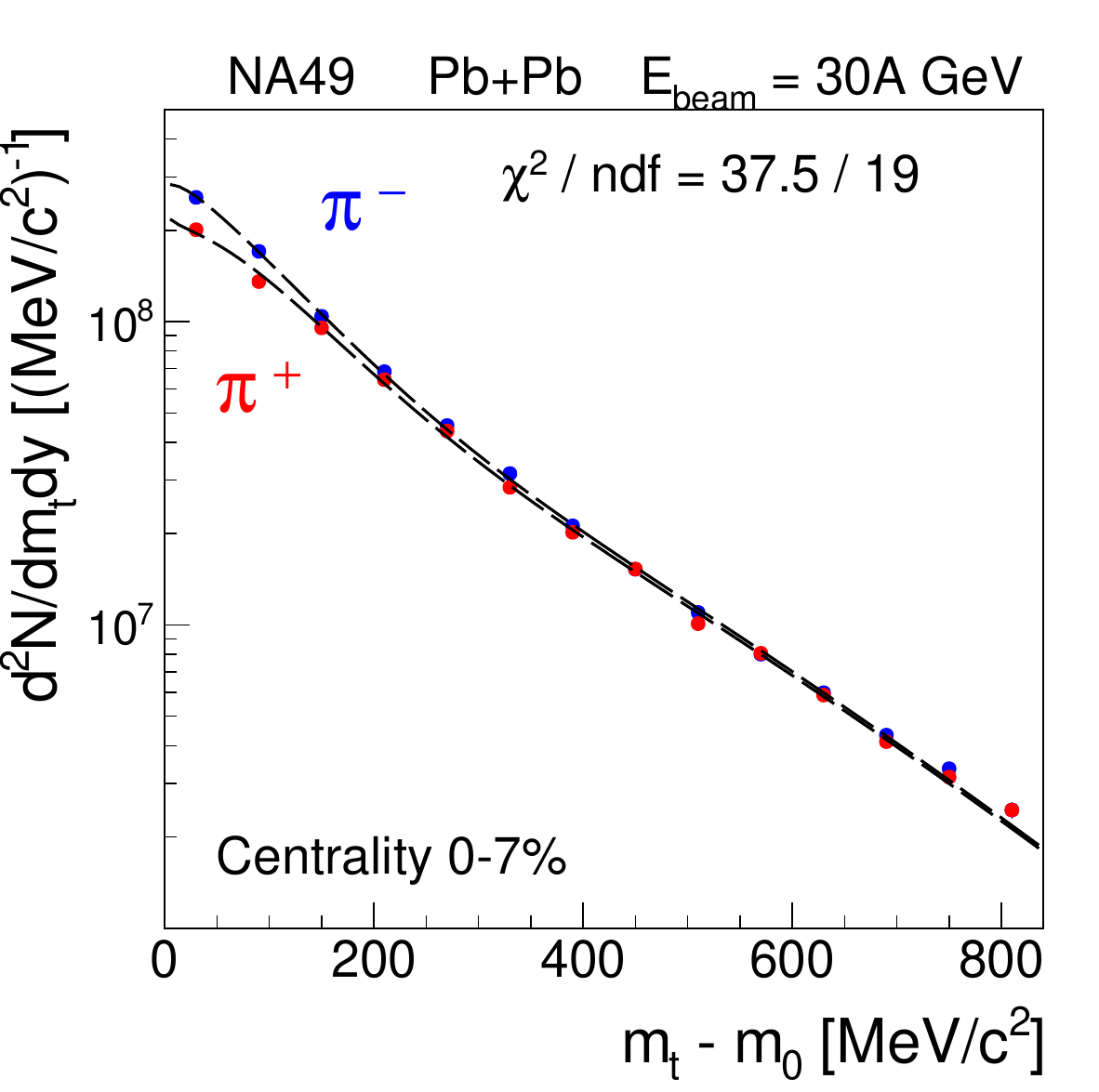}
	\caption{Coulomb-modified two-slope Boltzmann fits to E895 Au+Au and NA49 Pb+Pb data.
         The mid-rapidity transverse-mass spectra of $\pi^-$ (blue points) and $\pi^+$ (red points)
         measured by the AGS experiment E895 \cite{E-0895:2003oas} at 5\% centrality are shown for
         beam energies of 2, 4, 6, and 8\agev.  Solid curves correspond to a simultaneous fit of
         the $\pi^{+}$ and $\pi^{-}$ data points with Eq.~\eqref{eq-fit-mt-Veff}, assuming 3D expansion
         and setting $T_p$ to 187, 211, 216, and 229~MeV, respectively.
         The inserts show for 6 and 8\agev\ a close-up to the low-$\mt$ region
         with a second fit, done assuming 2D expansion, added as cyan line.  The NA49 pion spectra,
         taken from \cite{NA49:2007stj}, correspond to the 7\% most central events for the
         rapidity range of $0.0 < \ycm < 0.2$; here the fits were done with a 2D $V_{\mathrm{eff}}$,
         setting for 20 and 30\agev\ $T_p$ to 249 and 265~MeV, respectively \cite{NA49:2006}.
         All resulting fit parameters are listed in Table~\ref{tab-Vc-Worlddata}.
	}
    \label{fig-AGS_3Dcase}  
\end{figure}

To investigate the beam-energy dependence of the Coulomb effects, we also analyzed published mid-rapidity charged-pion spectra from the KaoS experiment at the SIS18 accelerator,
from the E895 experiment at the AGS, and from the NA49 experiment at the SPS. The KaoS data \cite{Wagner:1997sa} were measured in 1\agev\ Au+Au collisions for the 14 $\%$ most
central events at a fixed laboratory polar angle of $\theta_{\mathrm{lab}} = 44^{\circ} \pm 4^{\circ}$, corresponding to a narrow, but non-uniform center-of-mass acceptance bin
located around $\ycm \simeq 0.15$.  Figure \ref{fig-KAOS} shows a fit of the pion center-of-mass kinetic-energy distributions $d\sigma/dE^{\mathrm{kin} }_{\mathrm{c.m.}}$ with
an adapted\footnote{For this fit, the KaoS cross-section data was transformed from its kinetic-energy representation at fixed laboratory angle into a transverse-mass representation.}
version of Eq.~\eqref{eq-fit-mt-Veff}, setting $T_p = 120$ MeV.  The resulting Coulomb potential energy is $V_C = 15.6 \pm 2.2$ MeV and the energy-averaged
$\pi^{-}/\pi^{+}$ ratio is $2.20\pm0.10$. In case of E895, the published data correspond to the 5$\%$ most central Au+Au collisions measured at beam energies of 2, 4, 6, and 8\agev,
respectively \cite{E-0895:2003oas}.  Finally, the analyzed NA49 data were collected for the 7\% most central Pb+Pb collisions at 20 and 30\agev\ ~\cite{NA49:2007stj}.
When extending our analysis to higher bombarding energies (i.e.\ up to 30\agev), the assumption of a spherical fireball may not anymore be applicable.  At high energies,
the shape of the particle source close to mid-rapidity is usually assumed to be cylindrical, implying a 2D expansion rather than the 3D one expressed in Eq.~\eqref{eq-Veff-cases}.
To account for this effect, we consider both cases, 2D and 3D, in our analysis.  Figure \ref{fig-AGS_3Dcase} shows the fits done to the E895 and NA49 mid-rapidity charged-pion
$m_t$ spectra using Eq. \eqref{eq-fit-mt-Veff}, with $V_{\mathrm{eff}}$ corresponding to a 3D fireball up to 8\agev\ and 2D above; for 6 and 8\agev, both results are presented.
All fit parameters and their statistical uncertainties are listed in Table~\ref{tab-Vc-Worlddata}.  For the E895 data, the $T_2$ slopes were kept fixed to their published values
as the limited $m_t$ acceptance for positive pions of this experiment did not allow stable fits with both slopes varying freely.

\begin{table}
	\centering
	\begin{tabular}{ccccc}
		\hline
	    Experiment  & $E_\mathrm{beam}$ & $T_{1}$        & $T_{2}$        & $V_{C}$  \\ 
	                  &  [AGeV]   &  [MeV]         & [MeV]          & [MeV]  \\ \hline
	    3D~KaoS & 1         & 47.5 $\pm$ 0.5 &  85.0           & 15.6 $\pm$ 2.2    \\
	    ~~~~~~E895        & 2         & 59.3 $\pm$ 0.8 & 116.0           & 12.1 $\pm$ 1.3    \\
	    ~~~~~~E895        & 4         & 66.1 $\pm$ 1.2 & 135.0           & 8.3  $\pm$ 1.6    \\
	    ~~~~~~E895        & 6         & 65.4 $\pm$ 1.3 & 141.0           & 9.2  $\pm$ 1.7    \\
	    ~~~~~~E895        & 8         & 62.3 $\pm$ 1.3 & 141.0           & 7.4  $\pm$ 2.0    \\
            ~~~~~~NA49        & 20        & 58.4 $\pm$ 0.7 & 129.9 $\pm$ 2.5 & 7.0 $\pm$ 0.8     \\
            ~~~~~~NA49        & 30        & 55.5 $\pm$ 0.8 & 127.2 $\pm$ 1.7 & 4.2 $\pm$ 1.0     \\ \hline

            2D~E895  & 6         & 65.4 $\pm$ 1.3 & 141.0           & 5.3  $\pm$ 1.0    \\
            ~~~~~~E895         & 8         & 62.2 $\pm$ 1.3 & 141.0           & 4.5  $\pm$ 1.1    \\
	    ~~~~~~NA49        & 20        & 58.4 $\pm$ 0.7 & 129.9 $\pm$ 2.5 & 4.4 $\pm$ 0.5     \\
	    ~~~~~~NA49        & 30        & 55.5 $\pm$ 0.8 & 127.2 $\pm$ 1.7 & 2.4 $\pm$ 0.6     \\ \hline
	\end{tabular}
\vspace{0.5cm}
\caption[]{Parameters resulting from fits of Eq.~\eqref{eq-fit-mt-Veff} to KaoS, E895 Au+Au, and NA49 Pb+Pb data.
           The seven upper rows correspond to fits done assuming 3D expansion of the emitting source while
           in the four lower rows a 2D expansion is assumed.}
	\label{tab-Vc-Worlddata}
\end{table}

One can see from Figs.~\ref{fig-KAOS} and \ref{fig-AGS_3Dcase} that the Coulomb-modified fit function gives overall a good description of the available data sets.
Furthermore, when comparing all presented pion $m_t$ distributions, it appears that, going from low to high beam energy, the spacing between negative
and positive pion yields decreases.  This can be attributed to the gradual decrease of both the $\pi^{-}/\pi^{+}$ ratio and $V_\mathrm{eff}$ with increasing bombarding energy.  

\begin{figure}[htb]
	\centering
	\includegraphics[width=0.7\linewidth]{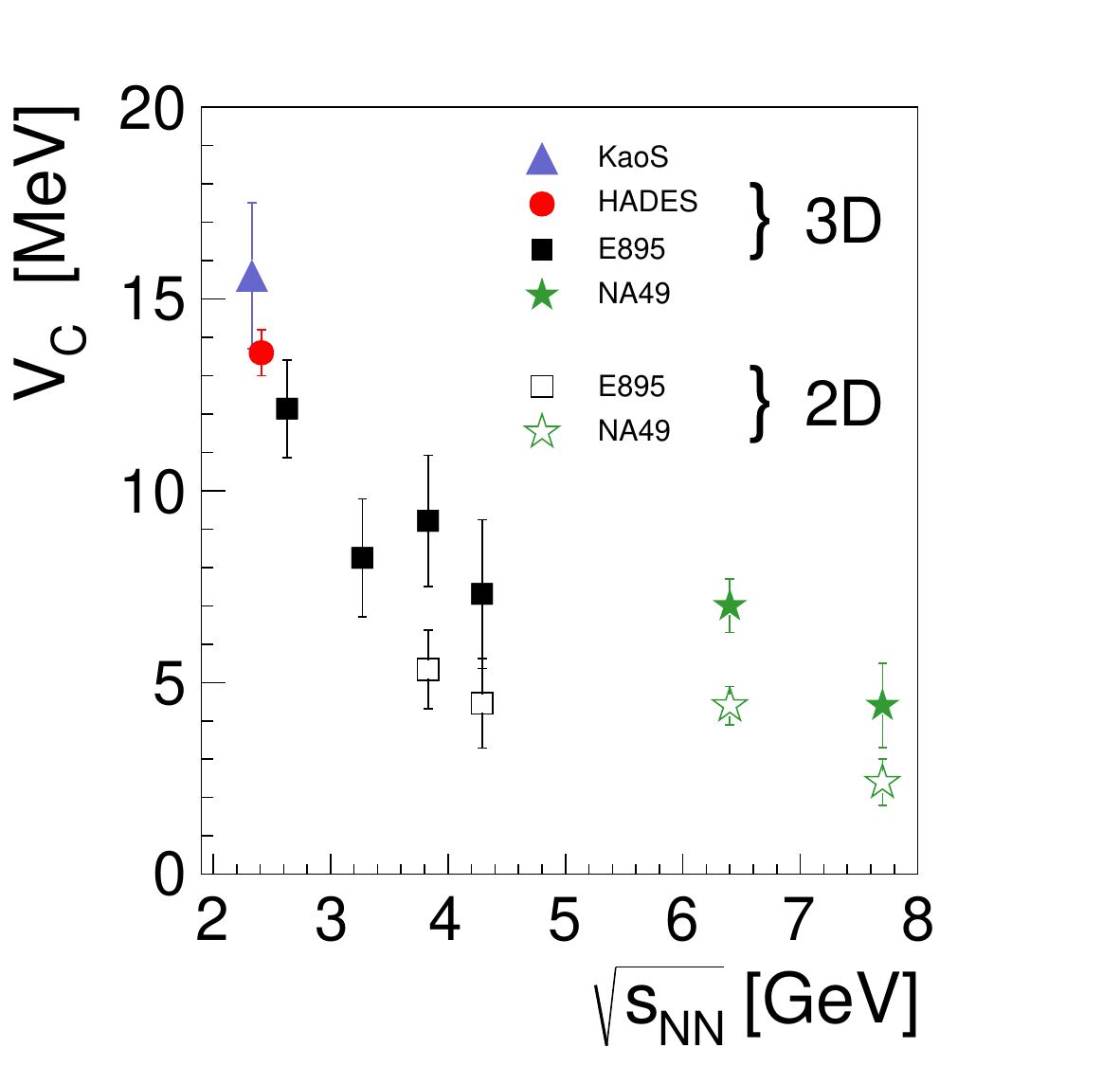}
	\caption{Coulomb potential energy $\Vc$ extracted from charged-pion mid-rapidity transverse-mass distributions in central Au+Au and Pb+Pb events,
          plotted as a function of the center-of-mass energy \sqrtsnn.  The data are from KaoS (blue triangle, Au+Au 0 -- 14\% centrality),
          HADES (red circle, Au+Au 0 -- 10\% centrality), E895 (black squares, Au+Au 0 - 5\% centrality), and NA49 (green stars, Pb+Pb 0 -- 7\% centrality).  
          The full (open) symbols correspond to fits assuming 3D (2D) expansion, respectively.}
    \label{fig-Vc-Ebeam}  
\end{figure}

Combining the results from all fits, the collision-energy dependence of the extracted Coulomb potential energy can be investigated for central events.  Although
the centrality selections of the various experiments are not identical, ranging between 0 -- 5\% and 0 -- 14\%, Fig. \ref{fig-Vc-Ebeam} shows that $V_C$ decreases
overall with increasing \sqrtsnn.  Such a trend is expected as, with rising energy, the center-of-mass momenta of the incident protons will be increasingly focused into
the longitudinal direction, diluting the electric field felt by the pions at mid-rapidity.  On the other hand, the limited accuracy of the AGS data precludes
to firmly decide whether the 3D or 2D expansion scenario matches better the general trend of the fitted Coulomb energy with \sqrtsnn.
Finally, note that an upper limit on $V_C$ can be estimated as given by the Coulomb potential energy of a charged sphere corresponding
to two fully overlapping gold nuclei, i.e.\ $V_C < 1.44 \, Z/R_{\circ} \simeq 33$~MeV, where $Z=2 \times 79$ and $R_{\circ} \simeq 7$~fm.

\section{Summary}

In this publication we have addressed the Coulomb interaction exerted on charged pions by the cumulated electric charge of the expanding fireball produced in relativistic heavy-ion collisions.
We have illustrated the observable effects with charged-pion spectra measured by the HADES spectrometer in \sqrtsnn\ =~2.4~GeV Au+Au reactions.  
 
Fitting the pion $m_t$ spectra with a Coulomb-modified two-slope Boltzmann distribution, we could reduce the systematic uncertainties arising in the determination of the
total pion yields.  Besides the $m_t$ inverse-slope parameters $T_1$ and $T_2$, those fits in addition deliver information on the average Coulomb potential energy experienced by
the produced particles.  Comparisons with results obtained by the former KaoS, E895, and NA49 experiments show that the latter quantity follows a smooth trend with center-of-mass energy.  

Relating the Coulomb potential energy to the spatial geometry of the expanding fireball, a freeze-out density can be calculated and, within the statistical hadronization model, constraints
on the baryochemical potential can be obtained.  Assuming volume emission of the pions, from either a hard-sphere or a Gaussian source, we deduced a range of baryon densities
of 0.11 - 0.12 $\rho_{0}$ and a baryochemical potential $\mu_{B}$ of 772 $\pm$ 10 MeV, consistent with values found in various SHM analyses of measured hadron yields.

The cumulative electric charge of the participant nucleons also affects the two-pion momentum correlations, as already pointed out in the HBT analysis of our Au+Au data
presented in \cite{Adamczewski-Musch:2018owj,Adamczewski-Musch:2019dlf}. We have shown that these second-order Coulomb manifestations are compatible with the effects
seen and quantified in the transverse-mass distributions of the charged pions. 


The present analysis of high-statistics charged-pion spectra has revealed the importance of low-momentum pions for a better understanding of the expansion phase of a heavy-ion collision.
These are pions moving slower than or with the expanding charged matter. This finding may encourage new fully dynamical calculations, e.g.\ transport simulations going beyond the
assumptions of the present analysis, to elucidate further the role of low-momentum pions for the dynamics governing the freeze-out of the fireball produced in relativistic heavy-ion collisions.

\begin{acknowledgements} 
\noindent
We thank Andreas Wagner for clarifying discussions and for providing the KaoS data points. 
We are furthermore grateful to Kai Gallmeister and Ulrich Mosel for elucidating discussions of the Coulomb effects in their GiBUU transport calculations.\\
\noindent
This work is in memory of our colleague Robert Greifenhagen.\\
\noindent
Financial support by the following agencies and grants is acknowledged:
SIP JUC Cracow, Cracow (Poland), 2017/26/M/ST2/00600; TU Darmstadt, Darmstadt (Germany), VH-NG-823, DFG GRK 2128, DFG CRC-TR 211, BMBF:05P18RDFC1;
Goethe-University, Frankfurt (Germany), BMBF:06FY9100I, BMBF:05P19RFFCA, GSI F\&E, HIC for FAIR (LOEWE); Goethe-University, Frankfurt (Germany) and TU Darmstadt, Darmstadt (Germany),
ExtreMe Matter Institute EMMI at GSI Darmstadt; NRNU MEPhI Moscow, Moscow (Russia), RFBR funding within the research project no. 18-02-00086,
Fundamental properties of elementary particles and cosmology No 0723-2020-0041; JLU Giessen, Giessen (Germany), BMBF:05P12RGGHM; IPN Orsay, Orsay Cedex (France), CNRS/IN2P3; NPI CAS,
Rez, Rez (Czech Republic), MSMT LTT17003, MSMT LM2018112, MSMT OP VVV CZ.02.1.01/0.0/0.0/18-046/0016066.

\end{acknowledgements}

\appendix
\section{Derivation of the Jacobian $J_{\mathrm{eff}}$}

In this appendix we derive the Jacobian $J_{\mathrm{eff}}$, expressed in Eq.~\eqref{eq-fit-jaceff}, introduced by the explicit momentum dependence of the effective Coulomb potential energy $V_{\mathrm{eff}}$ given by Eq.~\eqref{eq-Veff-cases}. To start, we recall that the Jacobian of the total energy transformation $E = E_0 \pm V_C$ caused by a constant Coulomb potential energy $V_C$ is given by

\begin{equation}
J = \left| \frac{\partial^3 p_0}{\partial^3 p} \right| = \left| \frac{p^2_0  \, \partial p_0}{p^2 \, \partial p}\right| = \frac{p_0 \, E_0 \, dE_0}{p \, E \ dE}= \frac{p_0 \, E_0}{p \, E} \; .
\label{eq-A1}
\end{equation}
The first equality follows from the spherical symmetry of the potential, i.e. by integration over polar and azimuthal angles, and the last one from the constancy of $V_C$ leading to $dE_0/dE = 1$.

If the constant $V_C$ is replaced by an energy-dependent effective potential $V_{\mathrm{eff}}(E)$, $E = E_0 + V_{\mathrm{eff}}(E)$ and the $dE_0/dE$ term changes into

\begin{equation}
\frac{dE_0}{dE} = \frac{d(E - V_{\mathrm{eff}}(E))}{dE} = 1 - \frac{dV_{\mathrm{eff}}(E)}{dE} \; .
\label{eq-A2}
\end{equation}

\noindent
As discussed in section~2.1, the expansion of the charged fireball leads to an attenuation of the average Coulomb potential acting on a charged pion of given energy $E = E(v)$, which we model, following Ref.~\cite{Barz:1998ce}, by the fraction of protons that have a lower velocity than the pion, $v_p < v_{\pi}$.  This results in the effective potential $V_{\mathrm{eff}}(E)$ expressed by Eq.~\eqref{eq-Veff-cases} for 2D and 3D expansions, respectively.  Setting the variable $x = \sqrt{(E_{\pi}/m_{\pi} - 1) \, m_p/T_p}$, we obtain by derivation 
\begin{equation}
\frac{dV_{\mathrm{eff}}}{dE} = \frac{dV_{\mathrm{eff}}}{dx} \; \frac{m_p}{m_{\pi} \, T_p} \, \frac{1}{2 x} \;\;\;= \;\;
\begin{cases}
 \;\; V_C \, \frac{m_p}{m_{\pi} \, T_p} \, e^{-x^2}  & \textrm{for 2D} ,\\
 \;\; Vc \, \frac{2 m_p}{\sqrt{\pi} \, m_{\pi} \, T_p} \, x \, e^{-x^2}  & \textrm{for 3D} , 
\end{cases}
\label{eq-A3}
\end{equation}
\noindent
from which the second Jacobian term in Eq.~\eqref{eq-fit-jaceff} follows.

\end{document}